\documentclass[12pt,a4paper]{article}
\pdfoutput=1
\usepackage{lineno}  
\usepackage{graphicx}  
\usepackage{xspace} 
\usepackage{color}
\usepackage{colortbl}
\usepackage{amsmath} 
\usepackage{ifthen} 

\usepackage{listings}
\usepackage{rotating}
\usepackage{multirow}
\usepackage{tabularx}
\newboolean{pdflatex}
\setboolean{pdflatex}{true} 
\newboolean{articletitles}
\setboolean{articletitles}{true} 

\newboolean{uprightparticles}
\setboolean{uprightparticles}{false} 

\usepackage{amssymb}
\usepackage{amsfonts}
\usepackage{upgreek} 

\usepackage{hyperref}    
\usepackage[all]{hypcap} 

\def\lhcb {LHCb\xspace}
\def\ux85 {UX85\xspace}

\ifthenelse{\boolean{uprightparticles}}%
{

 \def\PDelta      {\ensuremath{\Delta}\xspace}                 
 \def\PXi      {\ensuremath{\Xi}\xspace}                 
 \def\PLambda      {\ensuremath{\Lambda}\xspace}                 
 \def\PSigma      {\ensuremath{\Sigma}\xspace}                 
 \def\POmega      {\ensuremath{\Omega}\xspace}                 
 \def\PUpsilon      {\ensuremath{\Upsilon}\xspace}                 
 
 \def\PB      {\ensuremath{\mathrm{B}}\xspace}                 
                  
 \def\PD      {\ensuremath{\mathrm{D}}\xspace}

 \def\PK      {\ensuremath{\mathrm{K}}\xspace}

 \def\PW      {\ensuremath{\mathrm{W}}\xspace}

 \def\Pb      {\ensuremath{\mathrm{b}}\xspace}                 
 \def\Pc      {\ensuremath{\mathrm{c}}\xspace}

 \def\Pi      {\ensuremath{\mathrm{i}}\xspace}

}
{

 \mathchardef\PDelta="7101
 \mathchardef\PXi="7104
 \mathchardef\PLambda="7103
 \mathchardef\PSigma="7106
 \mathchardef\POmega="710A
 \mathchardef\PUpsilon="7107
                  
 \def\PB      {\ensuremath{B}\xspace}                 
                  
 \def\PD      {\ensuremath{D}\xspace}

 \def\PK      {\ensuremath{K}\xspace}

 \def\PW      {\ensuremath{W}\xspace}

 \def\Pb      {\ensuremath{b}\xspace}                 
 \def\Pc      {\ensuremath{c}\xspace}

 \def\Pi      {\ensuremath{i}\xspace}

}

\def\Wp     {\ensuremath{\PW^+}\xspace}
\def\Wm     {\ensuremath{\PW^-}\xspace}

\def\cquark    {\ensuremath{\Pc}\xspace}

\def\bquark    {\ensuremath{\Pb}\xspace}

\def\kaon  {\ensuremath{\PK}\xspace}
  \def\Kbar  {\kern 0.2em\overline{\kern -0.2em \PK}{}\xspace}

\def\Kz    {\ensuremath{\kaon^0}\xspace}
\def\Kzb   {\ensuremath{\Kbar^0}\xspace}
\def\KzKzb {\ensuremath{\Kz \kern -0.16em \Kzb}\xspace}
\def\Kp    {\ensuremath{\kaon^+}\xspace}
\def\Km    {\ensuremath{\kaon^-}\xspace}

\def\KpKm  {\ensuremath{\Kp \kern -0.16em \Km}\xspace}

  \def\Dbar    {\kern 0.2em\overline{\kern -0.2em \PD}{}\xspace}
\def\D       {\ensuremath{\PD}\xspace}

\def\Dz      {\ensuremath{\D^0}\xspace}
\def\Dzb     {\ensuremath{\Dbar^0}\xspace}
\def\DzDzb   {\ensuremath{\Dz {\kern -0.16em \Dzb}}\xspace}
\def\Dp      {\ensuremath{\D^+}\xspace}
\def\Dm      {\ensuremath{\D^-}\xspace}

\def\DpDm    {\ensuremath{\Dp {\kern -0.16em \Dm}}\xspace}

  \def\Bbar    {\kern 0.18em\overline{\kern -0.18em \PB}{}\xspace}

  \def\Y#1S{\ensuremath{\PUpsilon{(#1S)}}\xspace}

\def\AT#1     {\ensuremath{A_{\mathrm{T}}^{#1}}\xspace}           

\def\C#1      {\ensuremath{\mathcal{C}_{#1}}\xspace}                       
\def\Cp#1     {\ensuremath{\mathcal{C}_{#1}^{'}}\xspace}                    
\def\Ceff#1   {\ensuremath{\mathcal{C}_{#1}^{\mathrm{(eff)}}}\xspace}        
\def\Cpeff#1  {\ensuremath{\mathcal{C}_{#1}^{'\mathrm{(eff)}}}\xspace}       
\def\Ope#1    {\ensuremath{\mathcal{O}_{#1}}\xspace}                       
\def\Opep#1   {\ensuremath{\mathcal{O}_{#1}^{'}}\xspace}                    

\newcommand{\tev}{\ensuremath{\mathrm{\,Te\kern -0.1em V}}\xspace}
\newcommand{\gev}{\ensuremath{\mathrm{\,Ge\kern -0.1em V}}\xspace}
\newcommand{\mev}{\ensuremath{\mathrm{\,Me\kern -0.1em V}}\xspace}
\newcommand{\kev}{\ensuremath{\mathrm{\,ke\kern -0.1em V}}\xspace}
\newcommand{\ev}{\ensuremath{\mathrm{\,e\kern -0.1em V}}\xspace}
\newcommand{\gevc}{\ensuremath{{\mathrm{\,Ge\kern -0.1em V\!/}c}}\xspace}
\newcommand{\mevc}{\ensuremath{{\mathrm{\,Me\kern -0.1em V\!/}c}}\xspace}
\newcommand{\gevcc}{\ensuremath{{\mathrm{\,Ge\kern -0.1em V\!/}c^2}}\xspace}
\newcommand{\gevgevcccc}{\ensuremath{{\mathrm{\,Ge\kern -0.1em V^2\!/}c^4}}\xspace}
\newcommand{\mevcc}{\ensuremath{{\mathrm{\,Me\kern -0.1em V\!/}c^2}}\xspace}

\def\mum  {\ensuremath{\,\upmu\rm m}\xspace}

\def\pb {\ensuremath{\rm \,pb}\xspace}
\def\invpb {\ensuremath{\mbox{\,pb}^{-1}}\xspace}

\def\gsim{{~\raise.15em\hbox{$>$}\kern-.85em
          \lower.35em\hbox{$\sim$}~}\xspace}
\def\lsim{{~\raise.15em\hbox{$<$}\kern-.85em
          \lower.35em\hbox{$\sim$}~}\xspace}

\def\tell1  {TELL1\xspace}
\def\ukl1   {UKL1\xspace}

\newcommand{\zmumu}{$Z\rightarrow\mu\mu$\xspace}
\newcommand{\ztautau}{$Z\rightarrow\tau\tau$\xspace}
\newcommand{\wtaunu}{$W\rightarrow\tau\nu$\xspace}
\newcommand{\wmunu}{$W\rightarrow\mu\nu$\xspace}

\newcommand{\IP}{IP\xspace}
\newcommand{\ww}{\textit{W}\xspace}
\newcommand{\pseudow}{pseudo-\textit{W}\xspace}
\newcommand{\wplusxsec}{\mbox{$831\pm9\pm27\pm 29$\pb\xspace}}
\newcommand{\wminxsec}{\mbox{$656\pm8\pm19\pm 23$\pb\xspace}}
\newcommand{\wratio}{\mbox{$1.27\pm0.02\pm0.01$\xspace}}
\newcommand{\wzratio}{\mbox{$19.4\pm0.5\pm0.9$\xspace}}
\newcommand{\ptmu}{p^\mu_\mathrm{T}\xspace}
\newcommand{\etamu}{\eta^\mu\xspace}
\newcommand{\zxsec}{$76.7\pm1.7\pm3.3\pm2.7$\pb\xspace}
\newcommand{\sigmawp}{\sigma_{W^+\rightarrow\mu^+\nu}\xspace}
\newcommand{\sigmawm}{\sigma_{W^-\rightarrow\mu^-\bar{\nu}}\xspace}
\newcommand{\sigmaw}{\sigma_{W\rightarrow\mu\nu}\xspace}
\newcommand{\sigmaz}{\sigma_{Z\rightarrow\mu\mu}\xspace}
\newcommand{\DYNNLO}{{\sc Dynnlo}\xspace}
\newcommand{\ABKM}{ABKM09\xspace}
\newcommand{\CTEQl}{CTEQ6ll\xspace}
\newcommand{\MSTW}{MSTW08\xspace}
\newcommand{\CTEQM}{CTEQ6m\xspace}
\newcommand{\JR}{JR09\xspace}
\newcommand{\HERA}{HERA15\xspace}
\newcommand{\NNPDF}{NNPDF21\xspace}
\usepackage{cite}
\usepackage{mciteplus}
\begin{document}
\renewcommand{\thefootnote}{\fnsymbol{footnote}}
\setcounter{footnote}{1}


\begin{titlepage}
\pagenumbering{roman}

\vspace*{-1.5cm}
\centerline{\large EUROPEAN ORGANIZATION FOR NUCLEAR RESEARCH (CERN)}
\vspace*{1.5cm}
\hspace*{-0.5cm}
\begin{tabular*}{\linewidth}{lc@{\extracolsep{\fill}}r}
\ifthenelse{\boolean{pdflatex}}
{\vspace*{-2.7cm}\mbox{\!\!\!\includegraphics[width=.14\textwidth]{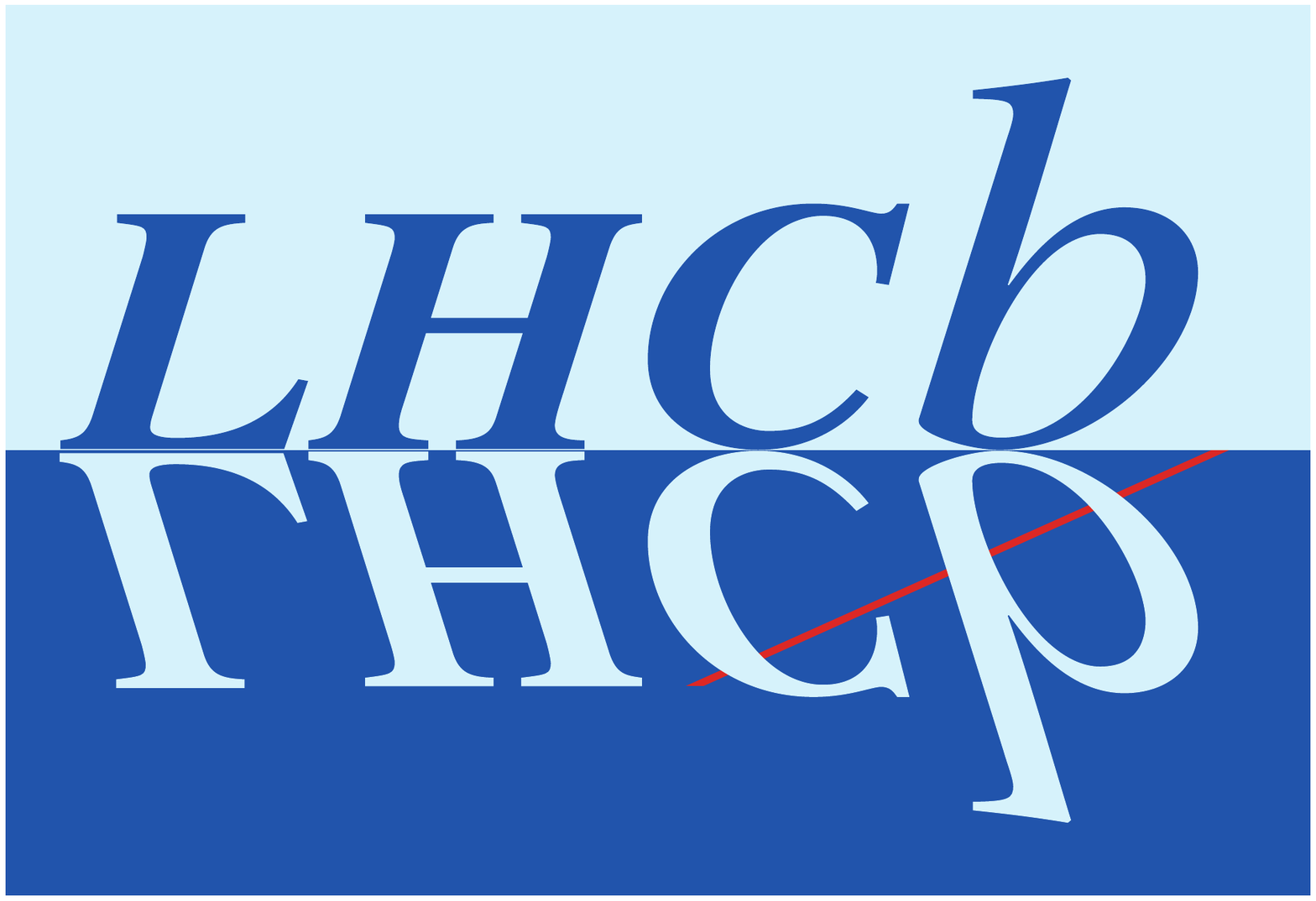}} & &}%
{\vspace*{-1.2cm}\mbox{\!\!\!\includegraphics[width=.12\textwidth]{figs/lhcb-logo.eps}} & &}%
\\
 & & CERN-PH-EP-2012-099 \\  
 & & LHCb-PAPER-2012-008 \\  
 & & May 24, 2012 \\ 
 & & \\
\end{tabular*}

\vspace*{3.0cm}

{\bf\boldmath\huge
\begin{center}
   Inclusive $W$ and $Z$ production in the forward region at $\sqrt{s} = 7$\tev 
\end{center}
}

\vspace*{1.5cm}

\begin{center}
The LHCb collaboration
\footnote{Authors are listed on the following pages.}
\end{center}

\vspace{\fill}

\begin{abstract}
  \noindent
Measurements of inclusive $W$ and $Z$  boson production cross-sections in $pp$ collisions 
at $\sqrt{s}=7$\tev using final states containing muons are presented.
The data sample corresponds to an integrated luminosity of $37$\invpb collected with the LHCb detector.
The $W$ and $Z$ bosons are reconstructed  from muons  with a transverse momentum 
above 20\gevc and pseudorapidity between $2.0$ and $4.5$, and, in the case of the $Z$ cross-section, a dimuon invariant mass     between $60$ and $120$\gevcc. 
The cross-sections are measured to be
\wplusxsec for $W^+$, 
\wminxsec for $W^-$ and
\zxsec for $Z$, where the first uncertainty is statistical,
the second is systematic and the third is due to the luminosity.
Differential cross-sections, $W$ and $Z$ cross-section ratios and the lepton charge asymmetry are
also measured in the same kinematic region.
The ratios are determined to be  $\sigmawp/\sigmawm= $\wratio\, and  $(\sigmawp+\sigmawm)/\sigmaz= $\wzratio.
The results are in general agreement with theoretical predictions, performed at
next-to-next-to-leading order in QCD using recently  calculated  parton distribution functions.
\end{abstract}

\vspace*{1.0cm}

\begin{center}
  Published in 
JHEP
Vol. 2012, Number 6 (2012), 58, DOI: 10.1007/JHEP06(2012)

\end{center}

\vspace{\fill}

\end{titlepage}


\newpage
\setcounter{page}{2}
\mbox{~}
\newpage

\centerline{\large\bf LHCb collaboration}
\begin{flushleft}
\small
R.~Aaij$^{38}$, 
C.~Abellan~Beteta$^{33,n}$, 
A.~Adametz$^{11}$, 
B.~Adeva$^{34}$, 
M.~Adinolfi$^{43}$, 
C.~Adrover$^{6}$, 
A.~Affolder$^{49}$, 
Z.~Ajaltouni$^{5}$, 
J.~Albrecht$^{35}$, 
F.~Alessio$^{35}$
% ...

{\footnotesize \it
% ...
$ ^{5}$Clermont Universit\'{e}, Universit\'{e} Blaise Pascal, CNRS/IN2P3, LPC, Clermont-Ferrand, France\\
$ ^{6}$CPPM, Aix-Marseille Universit\'{e}, CNRS/IN2P3, Marseille, France\\
% ...
$ ^{11}$Physikalisches Institut, Ruprecht-Karls-Universit\"{a}t Heidelberg, Heidelberg, Germany\\
% ...
$ ^{33}$Universitat de Barcelona, Barcelona, Spain\\
$ ^{34}$Universidad de Santiago de Compostela, Santiago de Compostela, Spain\\
$ ^{35}$European Organization for Nuclear Research (CERN), Geneva, Switzerland\\
% ...
$ ^{38}$Nikhef National Institute for Subatomic Physics, Amsterdam, The Netherlands\\
% ...
$ ^{43}$H.H. Wills Physics Laboratory, University of Bristol, Bristol, United Kingdom\\
% ...
$ ^{49}$Oliver Lodge Laboratory, University of Liverpool, Liverpool, United Kingdom\\
% ...
$ ^{n}$LIFAELS, La Salle, Universitat Ramon Llull, Barcelona, Spain\\
% ...
}
\end{flushleft}

\cleardoublepage


\renewcommand{\thefootnote}{\arabic{footnote}}
\setcounter{footnote}{0}



\pagestyle{plain} 
\setcounter{page}{1}
\pagenumbering{arabic}

 \section{Introduction}

The measurement of the production cross-sections for $W$ and $Z$ bosons  constitutes an important test of 
the Standard Model and
provides valuable input to constrain the proton parton density functions (PDFs).
Theoretical predictions are known to next-to-next-to-leading-order (NNLO) in perturbative quantum chromodynamics (pQCD). 
These calculations are in good agreement with recent measurements at the LHC from the \mbox{ATLAS}~\cite{atlas,atlas2}, and the CMS~\cite{cms,cms2} experiments as well as with the results from the $ p\bar{ p}$ collider experiments at the $\mathrm{S}\bar{\mathrm{p}\mathrm{pS}}$~\cite{ua1,ua2} and the Tevatron~\cite{cdf1,cdf2,cdf3,d01}.
The dominant theoretical uncertainty on the cross-sections arises from the present knowledge of the PDFs and the strong coupling constant.
The accuracy strongly depends on the pseudorapidity\footnote{The pseudorapidity $\eta$  is defined to be  $\eta = -\ln \tan(\theta/2)$, where the polar angle $\theta$ is measured with respect to the beam axis.}  range; consequently,
measurements by LHCb, which is fully instrumented in the forward region $2.0<\eta<5.0$,
 can provide input to constrain the PDFs, both for
pseudorapidities $\eta>2.5$ and in the region which is common to ATLAS and CMS, $2.0<\eta<2.5$.
Besides the determination of the $W$ and $Z$ boson cross-sections, the measurement of their ratios 
$R_{WZ}=(\sigmawp+\sigmawm)/\sigmaz$ and $R_W=\sigmawp/\sigmawm$ 
and of the $W$ production charge
asymmetry constitute important tests of the Standard Model, as experimental and theoretical uncertainties partially cancel.
The $W$ charge asymmetry is sensitive to the valence quark distribution in the proton~\cite{halzen} and provides complementary information to the results from deep-inelastic scattering cross-sections at HERA~\cite{h1zeus} as those data do not strongly constrain the ratio of $u$ over $d$ quarks at low Bjorken $x$, where $x$ is the proton momentum fraction carried by the quark. Measurements of $W$ and $Z$ boson production at LHCb have a sensitivity to values of $x$ as low as $1.7\times10^{-4}$ and will contribute significantly to the understanding of  PDFs at low $x$ and reasonably large four-momentum transfer $Q^2$, which corresponds to the squared mass of the $W$ or the $Z$ boson.

The measurements of the inclusive $W$ and $Z$ cross-sections\footnote{Throughout this paper $Z$ includes both the $Z$ and the virtual photon ($\gamma^\star$) contribution.} in $pp$ collisions at a centre-of-mass energy of $7$\tev,
 using final states containing muons, are presented in this paper.
The analysis is based on data taken by the LHCb experiment in $2010$ with  an integrated luminosity of $37$\invpb. 
The cross-sections are measured in a fiducial region corresponding to the
kinematic coverage of the LHCb detector, where the final state muons have a transverse momentum, $\ptmu$,  exceeding $20$\gevc and lie within the pseudorapidity range $2.0<\etamu<4.5$. 
This range is smaller than the LHCb acceptance in order to avoid edge effects for the acceptance.
In addition, the invariant mass of the muons from the $Z$ boson must be  in the range $60< M_{\mu\mu}<120$\gevcc.
Results are presented for the total cross-sections and  cross-section ratios.
Cross-sections  are also measured  
in bins of muon pseudorapidity for  $W$, and in
  bins of $Z$ rapidity ($y^Z$) for $Z$ production.  
Because of the presence of the neutrino, the production asymmetry between $W^+$ and $W^-$ cannot be reconstructed 
as a function of the boson rapidity. 
Instead it is measured as a function of the
experimentally accessible muon pseudorapidity, $\etamu$,  and referred to as the lepton charge asymmetry $A_\mu=(\sigmawp-\sigmawm)/(\sigmawp+\sigmawm)$.
To constrain the PDFs, it is useful to measure $A_µ$ for different $\ptmu$ thresholds.
 The data are compared to NNLO and NLO pQCD predictions with recent parametrisations for the PDFs. The signal efficiency and background contribution are  mostly derived from data.

The remainder of the paper is organised as follows. Section 2 describes the LHCb detector and the Monte Carlo samples. Section 3 describes the selection of the $W$ and $Z$ candidates, the backgrounds, the determination of the purity and the signal efficiencies. The measurement of the cross-sections as well as the systematic uncertainties are discussed in Sect. 4. The results are presented in Sect. 5 and conclusions in Sect. 6.

 
\section{LHCb detector and Monte Carlo samples}

The \lhcb detector~\cite{lhcb_detector} is a single-arm forward
spectrometer covering the pseudorapidity range $2<\eta<5$, designed
for the study of particles containing \bquark or \cquark quarks. The
detector includes a high precision tracking system consisting of a
silicon-strip vertex detector (VELO) surrounding the $pp$ interaction region,
a large-area silicon-strip detector (TT) located upstream of a dipole
magnet with a bending power of about $4{\rm\,Tm}$, and three stations
of silicon-strip detectors (IT) and straw drift-tubes (OT) placed
downstream. The combined tracking system has a momentum resolution
$\Delta p/p$ that varies from 0.4\% at 5\gevc to 0.6\% at 100\gevc,
and an impact parameter (IP) resolution of 20\mum for tracks with high
transverse momentum. Charged hadrons are identified using two
ring-imaging Cherenkov detectors. Photon, electron and hadron
candidates are identified by a calorimeter system consisting of
scintillating-pad and pre-shower detectors, an electromagnetic
calorimeter and a hadronic calorimeter. Muons are identified by a muon
system composed of alternating layers of iron and multiwire
proportional chambers. The trigger consists of a hardware stage, based
on information from the calorimeter and muon systems, followed by a
software stage which applies a full event reconstruction.
To avoid the possibility that a few events with high occupancy dominate the  CPU time of the software trigger, a set of global event cuts is applied on the hit multiplicities of most subdetectors used in the pattern recognition algorithms.

Several Monte Carlo (MC) simulated samples are used to develop the event selection, estimate the backgrounds, cross-check the efficiencies and to account for the effect of the underlying event. 
The {\sc Pythia} $6.4$~\cite{pythia} generator, configured
as described in Ref.~\cite{config}, with the \CTEQl~\cite{cteq} parametrisation for the PDFs is used to simulate the processes  
\zmumu, \ztautau, \wmunu  and \wtaunu.
The  hard partonic interaction is calculated in leading order pQCD and
higher order QCD radiation is modelled using initial and final state parton showers in the leading log approximation~\cite{Bengtsson:1987rw}. 
The fragmentation into hadrons is simulated in {\sc Pythia} by the Lund string model~\cite{lund}. 
All generated events are passed through a {\sc Geant4}~\cite{geant} based detector simulation, the trigger emulation and the event reconstruction chain of the LHCb experiment.

Samples of \wmunu and \zmumu simulated events with one muon in the LHCb acceptance  have been reweighted to  reproduce the NNLO $\ptmu$ distribution. These samples are referred to as $W$-MC and $Z$-MC, respectively. 
In the first step a  correction factor is calculated as a function of the generated muon transverse momentum by determining the ratio of the generated $\ptmu$ spectrum, as simulated by the {\sc Powheg}~\cite{powheg,*powheg1,*powheg2} generator at NLO, to the generated $\ptmu$ spectrum from {\sc Pythia}. In the second step the events are reweighted with a factor given by the ratio between the NNLO and NLO prediction as calculated with \DYNNLO~\cite{dynnlo}. This factor is calculated as a function of the rapidity of the boson.
As an alternative,   {\sc Pythia} samples have been reweighted to reproduce the $\ptmu$ distribution as calculated with {\sc Resbos}~\cite{resbos,*resbos1,*resbos2}.
{\sc Resbos} includes a NLO calculation plus next-to-next-to-leading-log  resummation of QCD effects at low transverse momentum.

 {\boldmath \section{Selection of $W$ and $Z$ events}}

\subsection{Muon reconstruction and identification}
Events with high transverse momentum muons are selected  using a single muon trigger with a threshold of  $\ptmu>10$\gevc.
Tracks are reconstructed starting from the VELO, within which particle trajectories are approximately straight, since
the detector is located upstream of the magnet.
  Candidate tracks are extrapolated
to the other side of the magnet and a search is made for compatible
hits in the IT and OT sub-detectors.  
An alternative strategy searches for track segments in  both the VELO and 
IT/OT detectors and extrapolates each to the bending plane of the magnet,
where they are matched.
Once VELO and IT/OT hits have been combined, 
an estimate of the track momentum is available and the full trajectory
can be defined.
  Finally, hits in the TT sub-detector are added if consistent with the 
candidate tracks.  Thus, the presence of TT hits can be considered as an
independent confirmation of the validity of the track.

Muons are identified by extrapolating the tracks and searching for
compatible hits in the four most downstream muon stations.  
For the high momentum muons that concern this analysis,
hits must be found in all four muon stations.  In total, the muon candidate
must have passed through over $20$ hadronic interaction lengths of material.

\begin{figure}[!htb]
  \begin{center} 
    \includegraphics[width=0.68\textwidth]{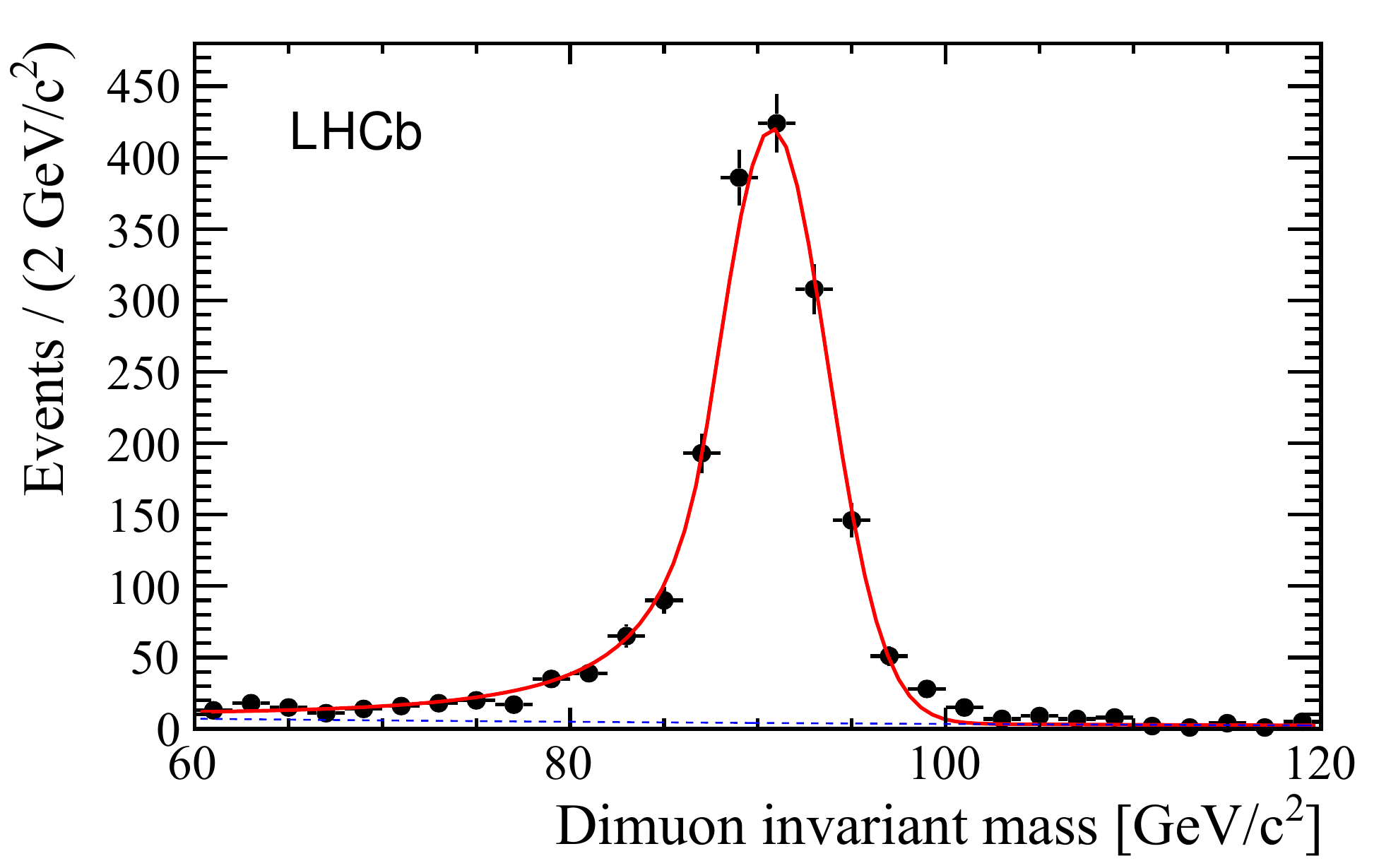}
    \caption{Invariant mass of the selected muon pairs. 
The fitted distribution to the data is shown as a solid line and
the contribution from background and off-resonance Drell-Yan production as a dashed line.
}
    \label{fig:zpeak}
  \end{center}
\end{figure}

{\boldmath \subsection{Selection of  \zmumu candidates}}
\label{sec:zsel}
Candidate \zmumu events are selected by requiring a pair of  well 
reconstructed tracks identified as muons;
 the invariant mass of the two muons  must be  in the range
 \mbox{$60< M_{\mu\mu}<120$\gevcc}.
Each muon track must have  $\ptmu>20$\gevc and lie in the range $2.0<\etamu<4.5$. 
The relative uncertainty on the momentum measurement is required to be  less than $10$\% and 
the probability for the $\chi^2/$ndf for the track fit  larger than $0.1$\%, where ndf is the number of degrees of freedom. 
In total, $1966$ $Z$ candidates are selected; their mass distribution is shown in  Fig.~\ref{fig:zpeak}. 
The data are not corrected for initial  or final state radiation. 
 A Crystal Ball~\cite{crystal} function for the $Z$ peak, and an exponential distribution for both the off-resonance Drell-Yan ($\gamma^\star$) production and the small  background contribution are fitted to the distribution.
The fitted mass $90.7\pm0.1$\gevcc and width $3.0\pm0.1$\gevcc, where the uncertainties are statistical, are consistent with expectation from simulation.

\vspace{1em}
{\boldmath \subsection{\zmumu event yield}}
The background contribution to the \zmumu analysis is very low. Five different sources are investigated.
\begin{enumerate}
\item Decays from \ztautau  contribute, if both taus decay leptonically to muons and neutrinos. The tau background is estimated from simulation, with the $Z$ cross-section fixed to the cross-section measured in this analysis, to contribute $0.6\pm0.1$ events to the total sample.
\item  Decays of  heavy flavour hadrons  contribute to the  background if they decay semileptonically (``heavy flavour'' background).
The contribution is estimated from a sample, which is enriched in background.
``Non-isolated'' muons  are selected with $\ptmu>15$\gevc and  $M_{\mu\mu}>40$\gevcc
and the scalar sum of the  transverse momenta of all tracks in a cone of half angle $0.5$ in $\eta\mbox{-}\phi$ around the muons  larger than $4$\gevc; here $\phi$ is the azimuthal angle measured in radians. 
A fit to the  invariant mass distribution at low masses is then used to estimate the background contribution in the  $Z$ mass region.
 The heavy flavour contribution is estimated to be  $3.5\pm0.8$ events.
\item Pions or kaons may be misidentified as muons if they decay in flight (``decay-in-flight'' background) or 
if they travel through the calorimeters and are identified by the muon chambers (``punch-through'' background).
This background  should contribute equally in same-sign and opposite-sign combinations of the muon pair. 
No event is found in the $Z$ selection with both tracks having the same charge. 
The contribution from muon misidentification is estimated to be less than one event.
\item  $W$ pair production contributes to the sample if both $W$ bosons decay  to a muon and a neutrino. This contribution corresponds to $0.2\pm 0.1$ events as estimated with {\sc Pythia} MC simulation.
\item Decays of top quark pairs  may contribute if both top quarks decay semileptonically. {\sc Pythia} MC simulation predicts a contribution of $0.5\pm 0.2$ events.
\end{enumerate}

The total background contribution in the $Z$ sample in the range $60\mbox{--}120$\gevcc amounts to $4.8\pm1.0$ events. This corresponds to a purity $\rho^Z=0.997\pm0.001$.   The purity is defined as the ratio of signal to candidate events. 
No significant dependence on the boson rapidity is observed.

{\boldmath\subsection{Selection of \wmunu candidates}}
\label{sec:wsel}
In leading order QCD, \wmunu events are characterised by a single high transverse
momentum muon that is not associated with other activity in the event. 
As only the muon can be reconstructed in LHCb, the background contribution is  larger for the $W$ 
than for the $Z$ candidates.
Therefore, more stringent requirements are placed on the track quality of the muon
and additional criteria are imposed in order to select $W$ candidates.

The optimisation of the $W$ selection and the evaluation of the selection efficiency
make use of a ``pseudo-$W$'' control sample obtained from the previously described
$Z$ selection, where each of the muons is masked in turn,
in order to mimic the presence of a neutrino and fake a \wmunu decay.
Excellent agreement is observed for all variables of interest between
pseudo-$W$ and $W$ simulated samples with the exception of those that have an explicit
dependence on the transverse momentum of the muon, as the underlying momentum distribution differs 
for muons from $Z$ and $W$.

The identification of \wmunu candidate events 
starts by requiring a well reconstructed
track which is identified as a muon.  The track must have a transverse momentum in the range
$20<\ptmu<70$\gevc within a pseudorapidity range $2.0<\etamu<4.5$.
The relative
error on the momentum measurement must be less than $10$\%, 
the probability for  the  $\chi^2/$ndf 
of the track fit must be greater than $1$\%, 
and there must be  TT hits associated to the track.
The last requirement reduces the number  of
combinations of VELO and IT/OT information that
have been incorrectly combined to form tracks.

\begin{figure}[thb!]
\begin{center}
\includegraphics[width=0.68\textwidth]{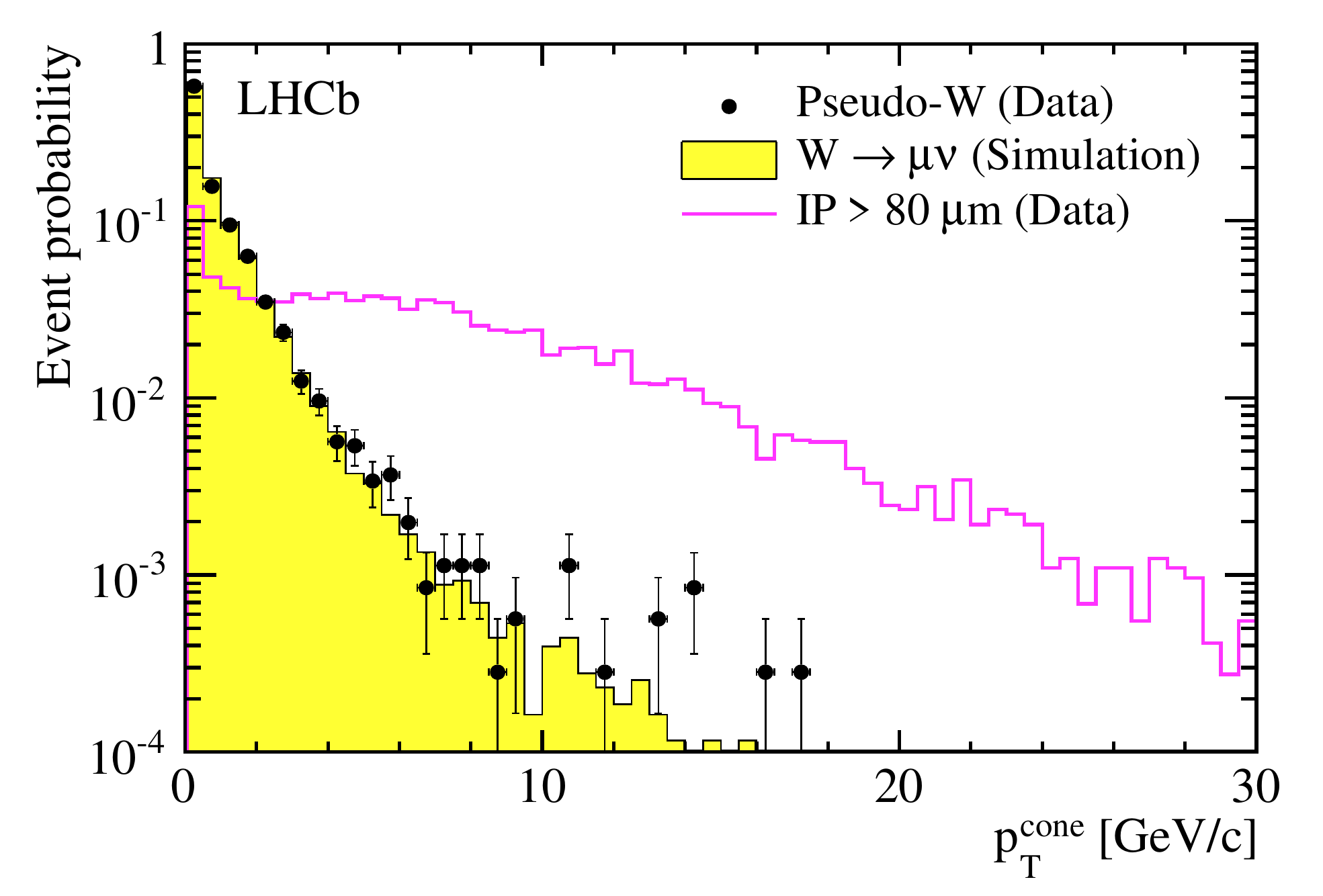}
\includegraphics[width=0.68\textwidth]{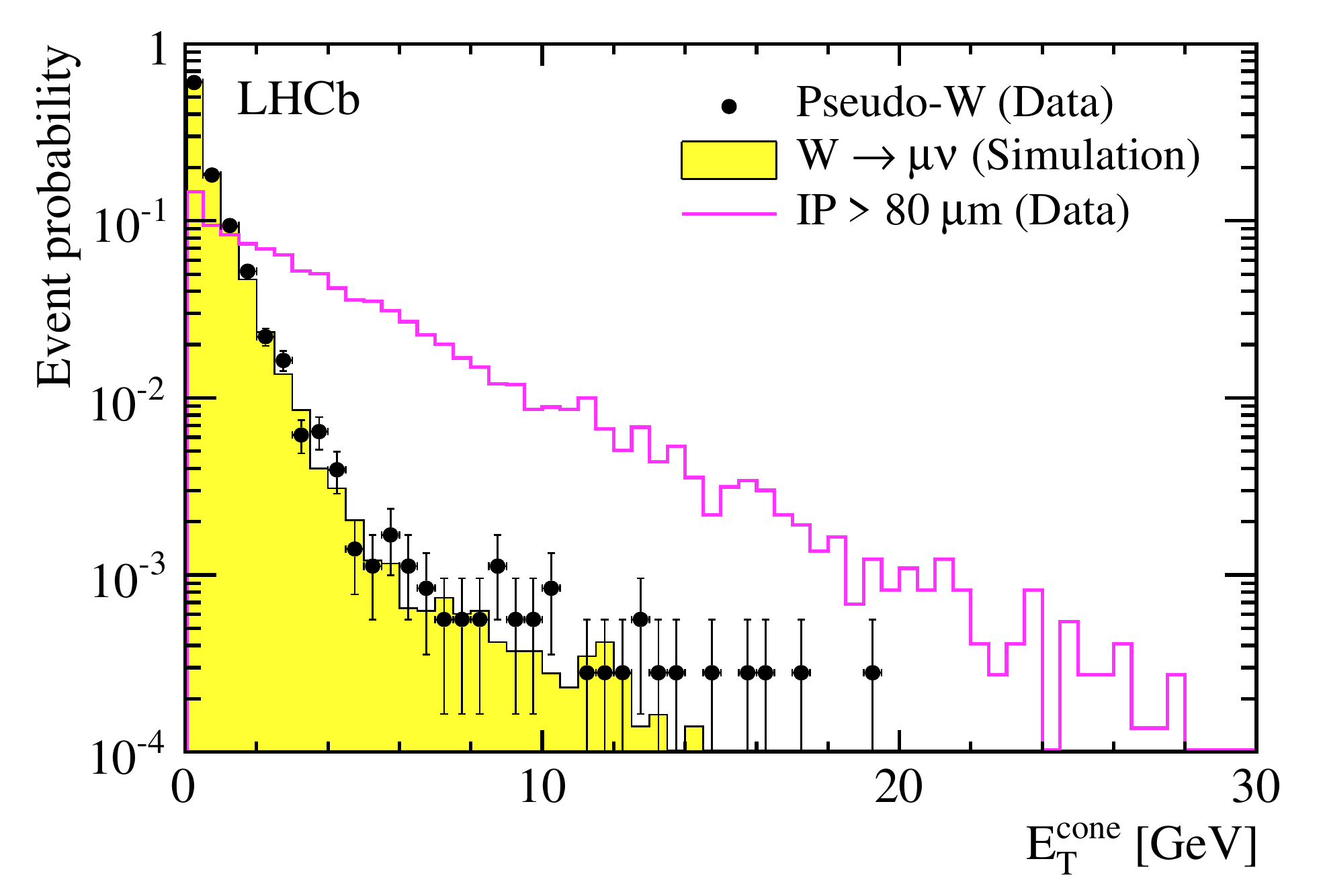}
\end{center}
\caption{
Distributions for $p_\mathrm{T}^\mathrm{cone}$ (top) and $E_\mathrm{T}^\mathrm{cone}$ (bottom).
The points are for muons from pseudo-$W$ data, the yellow  (shaded) histograms are for $W$-MC simulation,
while the open histograms are for muons from QCD background with \IP$>80$\mum from data.
All distributions are normalised to unity.
}
\label{fig:conehf}
\end{figure}

To suppress background from \zmumu decays, it is required that any other 
identified muon in the event has a transverse momentum below 2\gevc.  
This removes the
events where both muons have entered the LHCb acceptance.

Identified muons can originate from background processes of heavy flavour decays,
or misidentification of pions and kaons due to decay-in-flight or punch-through (``QCD background'').
In all such cases, the identified muon is usually produced in the same direction as the
other fragmentation products, in contrast to muons from $W$ decays which tend to be isolated.
The isolation of the muon is described using the charged transverse momentum, $p_\mathrm{T}^\mathrm{cone}$, and neutral
transverse energy, $E_\mathrm{T}^\mathrm{cone}$, in a cone around the candidate muon.
The quantity $p_\mathrm{T}^\mathrm{cone}$ is defined as the scalar sum of the transverse momentum of all tracks,
excluding the candidate muon, satisfying 
$\sqrt{(\Delta\phi)^2+(\Delta\etamu)^2}<0.5$, 
where $\Delta\phi$ and $\Delta\etamu$ are the differences in $\phi$ and $\eta$ between the
muon candidate and the track.
The quantity $E_\mathrm{T}^\mathrm{cone}$ is defined in a similar way, but summing
the transverse energy of all electromagnetic calorimeter deposits not associated with tracks.
The distributions for $p_\mathrm{T}^\mathrm{cone}$ and $E_\mathrm{T}^\mathrm{cone}$
are shown in Fig.~\ref{fig:conehf} for \pseudow data,  $W$-MC 
and muons  with $p_\mathrm{T}^\mu>20$\gevc  and an IP  larger than $80$\mum. 
The \IP of the muon is
defined as the distance of closest approach to the primary vertex
calculated from the other tracks in the event excluding the muon candidate.
The sample with high \IP  is 
enriched with muons from decays of heavy flavour hadrons, showing the typical shape of  QCD background.
There is  agreement between \pseudow data and $W$-MC, while the shape
for the heavy flavour events is quite different.  
To suppress QCD  background, 
it is required that $p_\mathrm{T}^\mathrm{cone}<2$\gevc and $E_\mathrm{T}^\mathrm{cone}<2$\gev.

\begin{figure}[!thb]
  \begin{center} 
    \includegraphics[width=0.68\textwidth]{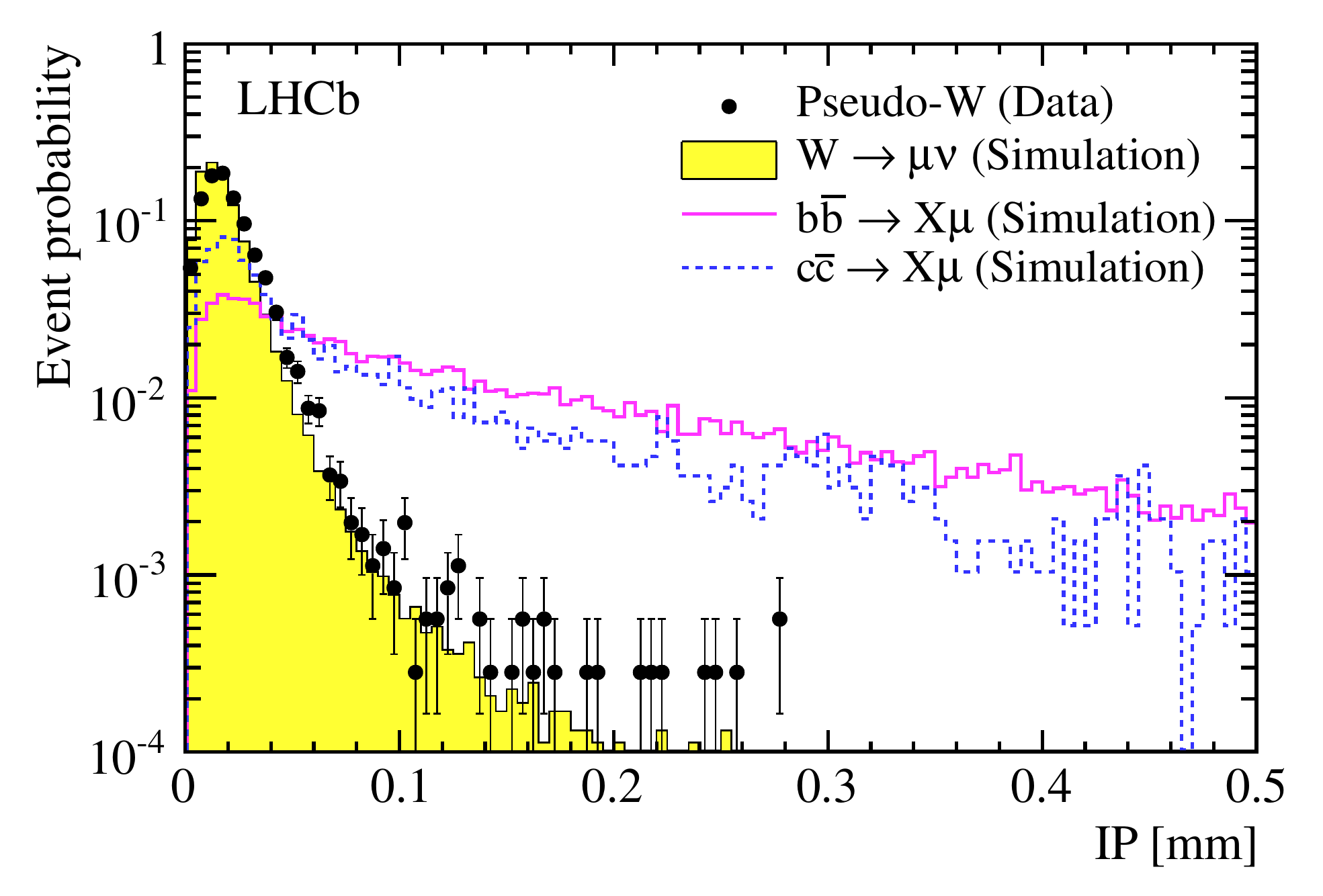}
    \caption{ Muon IP distribution for
\pseudow events as points, $W$-MC as a yellow   (shaded) histogram, and 
muons from simulated semi-leptonic
decays of hadrons containing a $b$ quark in the full open histogram or  
a $c$ quark in the dashed open histogram.
All distributions are normalised to unity.
}
    \label{fig:life}
  \end{center}
  \begin{center} 
    \includegraphics[width=0.68\textwidth]{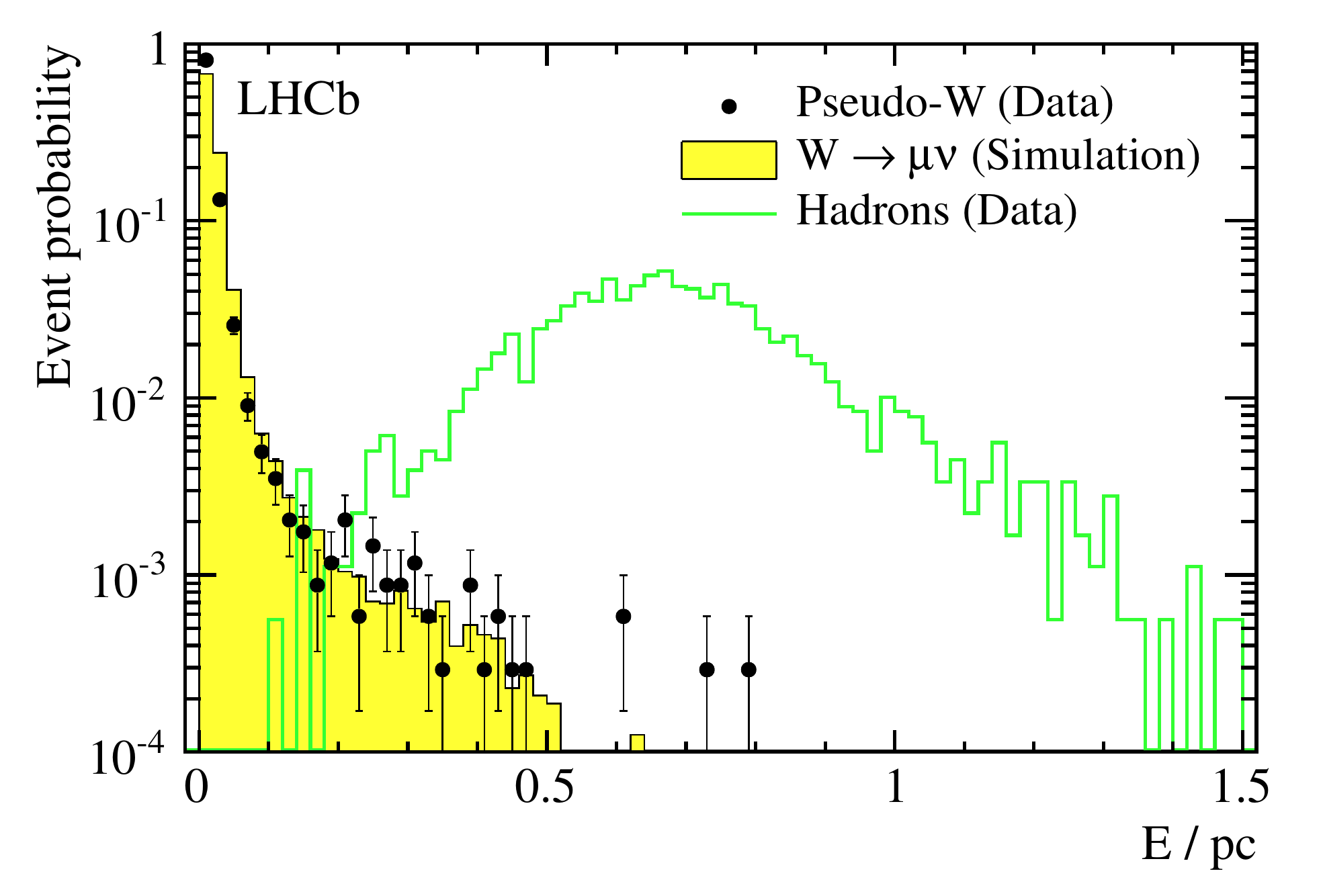}
    \caption{$E/pc$ for \pseudow events as points, $W$-MC as a yellow (shaded) histogram,
and for hadrons from randomly triggered events in the open histogram. The  energy $E$ is the
 sum of the energies  in the electromagnetic and hadronic calorimeter associated 
with the particle.
All distributions are normalised to unity.
}
    \label{fig:punch}
  \end{center}
\end{figure}

Muons originating from semi-leptonic decays of heavy flavour hadrons can be further suppressed by a cut on the \IP.
Due to the lifetimes of the $B$ and $D$ mesons, these muons do not originate from the 
primary $pp$ interaction.
The \IP distribution is shown
in Fig.~\ref{fig:life} for \pseudow events, $W$-MC,
and simulated semi-leptonic decays of hadrons containing a $b$ or $c$ quark.
The \pseudow events and $W$-MC are in  agreement and
peak at low values of \IP, in contrast to the heavy flavour background.  For the $W$ candidate selection it is required that
\IP$<40$\mum.
This cut also removes a large fraction of the background from \wtaunu and \ztautau decays.

Pions and kaons that punch-through to the muon chambers can be distinguished from true muons
as they leave substantial energy deposits in the calorimeters.
Figure~\ref{fig:punch} shows the summed energy, $E$, in the electromagnetic and hadronic calorimeter associated 
with the particle, divided by the track momentum, $p$, for pseudo-$W$ events,
$W$-MC, and hadrons with $p_\mathrm{T}>20$\gevc in randomly
triggered events.  By requiring $E/pc<0.04$ the punch-through contamination can be reduced
to a negligible level.  
The disagreement between pseudo-$W$ data and simulated $W$-MC in Fig.~\ref{fig:punch} is 
caused by the different 
 underlying momentum distribution for muons from $W$ and $Z$.

\vspace{1em}
{\boldmath \subsection{\wmunu event yield}}\label{sec:yield}
After the $W$ selection requirements are imposed $14\,660$ $W^+$ and $11\,618$ $W^-$ candidate events are observed.
The \wmunu signal yield has been determined by fitting the $\ptmu$ spectra
of positive and negative muons  in data, to template shapes for
signal and backgrounds in five bins of $\etamu$.  
The fit is performed with the following sources for signal and background with the shapes and normalisations as 
described below.  
\begin{enumerate}

\item
The \wmunu signal template is obtained using the $W$-MC.
The normalisation is left free to vary in each bin of $\etamu$ and
for each charge. 

\item
The shape of the  template of the largest background, \zmumu, is taken from the $Z$-MC. 
The normalisation is fixed from data by counting
the number of $Z$ events, scaled by the ratio of events with one muon in
the LHCb acceptance to events with both muons in the acceptance, as determined from $Z$-MC.
The ratio is  corrected for 
the different reconstruction and selection efficiencies for $W$ and $Z$ as derived from data.
This gives an expectation of $2435\pm101$ background events  
 (($9.3\pm0.4$)\% of the total sample)
in good agreement with  $2335\pm25$ events  found from simulation.

\item
The shape of the \wtaunu and \ztautau templates are taken from 
{\sc Pythia}. The \ztautau template is scaled according to the observed number of $Z$ events.
These $\tau$ backgrounds 
constitute $2.7$\% of the total sample.

\item
The heavy flavour template is obtained from data by requiring that the muon
is not consistent with originating from the primary vertex  (\IP$>80$\mum).
 The normalisation is determined from data applying all requirements
 except for the impact parameter and fitting the resulting \IP distribution  
 to the two templates shown in Fig.~\ref{fig:life}:
 the pseudo-$W$ data to describe the signal, and the
simulated heavy flavour events to describe the background.  The heavy flavour contribution
is estimated to be ($0.4\pm0.2$)\% of the total sample.

\item 
The punch-through contribution from kaons and pions is 
largely suppressed by the requirement on $E/pc$. 
The $E/pc$ distribution in Fig.~\ref{fig:punch} is fitted to  pseudo-$W$ data for the signal, and a Gaussian for the  punch-through, in order to estimate the punch-through contribution.  
 This is found to be negligible
($0.02 \pm 0.01$)\% of the total sample, and also has a shape very similar to the decay-in-flight
component. Hence, this component is not considered when determining the signal yield.
\item
The decay-in-flight shape is found from data in a two-step procedure using all events 
selected throughout $2010$ by any trigger requirement.
First,  tracks with a transverse momentum between $20$ and $70$\gevc are taken to describe the $p_\mathrm{T}$ spectrum of hadrons;
 tracks that fired a muon trigger are excluded from the sample.
Second, this spectrum is weighted according to the probability for a hadron
to decay-in-flight.  This probability is defined as the
fraction of tracks identified as muons in randomly triggered events
and is parametrised as a function of the momentum, $p$, by a function of
the form
\begin{equation}
 1-e^{-\alpha/p},
\label{equ:fit}
\end{equation}
 as would be expected for a particle whose mean
lifetime in the laboratory
frame scales with its boost.
Consistent values for $\alpha$ are found in each pseudorapidity bin and are
in agreement with a calculation of the decay probability based on the
mean lifetimes for charged pions and kaons, and the
distance to the electromagnetic calorimeter before which the hadron must have decayed.
The average of the determinations in each pseudorapidity
bin defines the central value for $\alpha$.
The relative normalisation of positively to negatively charged
tracks in each bin of pseudorapidity is fixed to that observed in randomly triggered events, 
but the overall normalisation in each bin of pseudorapidity is left free.
\end{enumerate}
\begin{figure}[!htb]
\begin{center}
\includegraphics[width=0.49\textwidth]{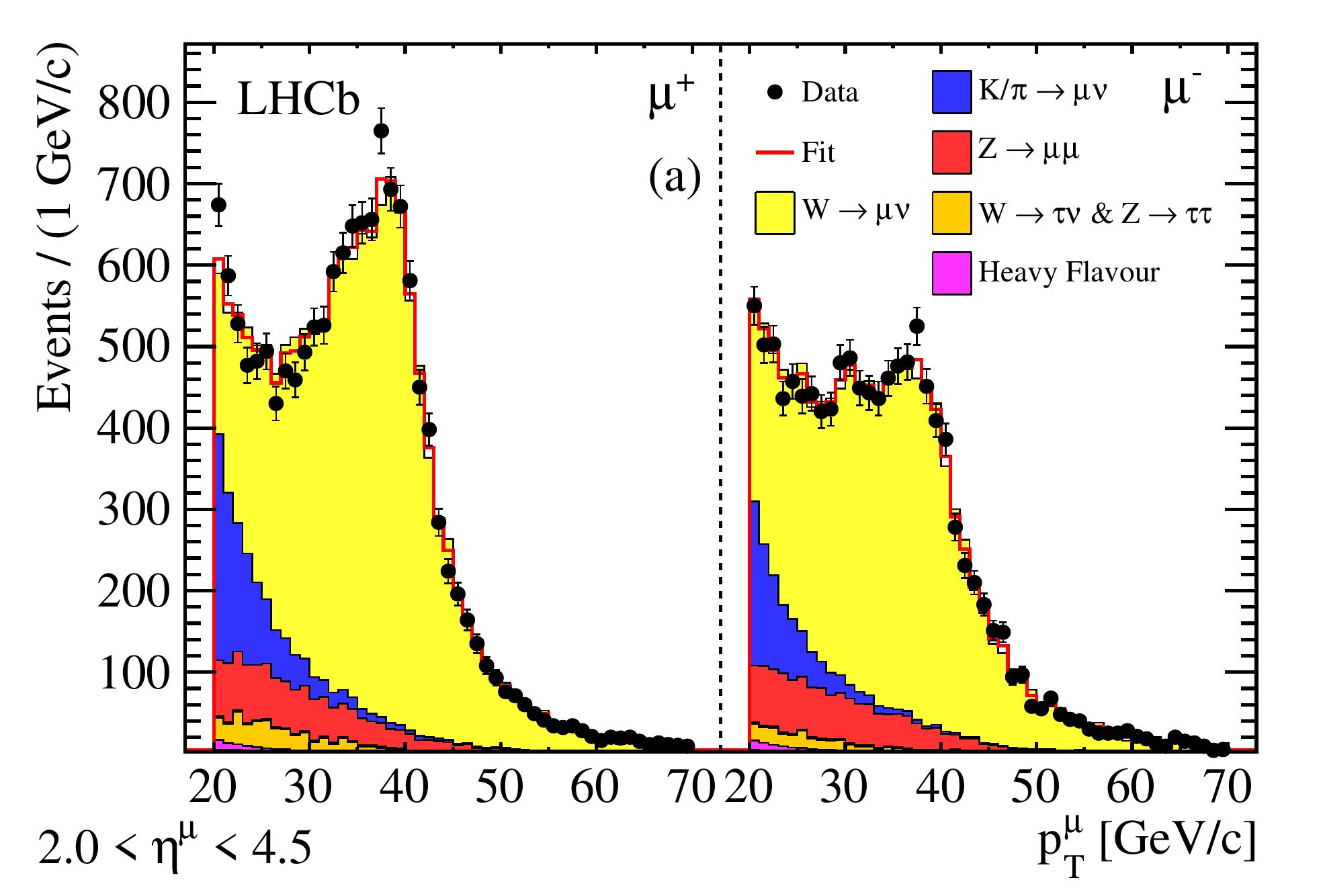}
\includegraphics[width=0.49\textwidth]{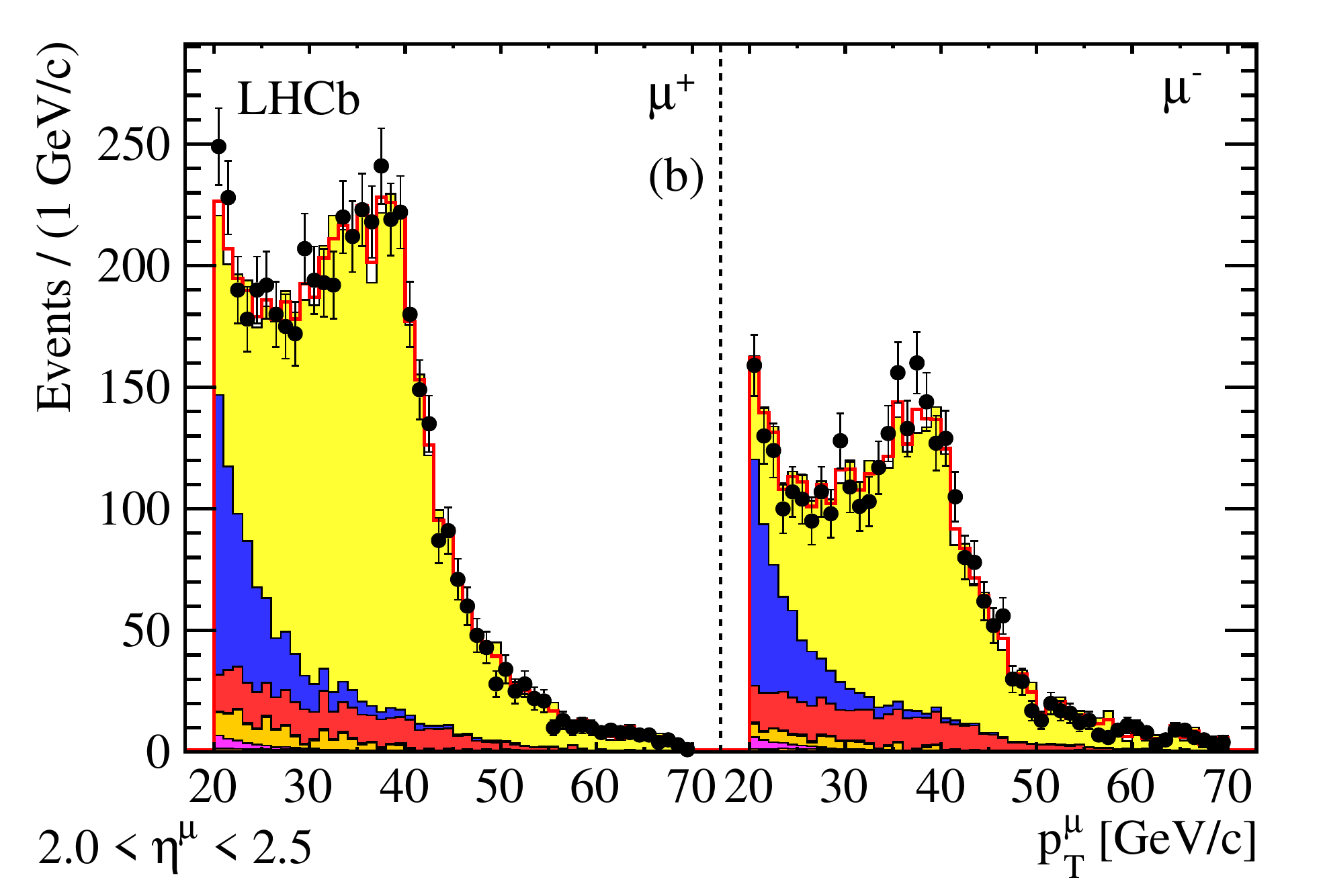}
\includegraphics[width=0.49\textwidth]{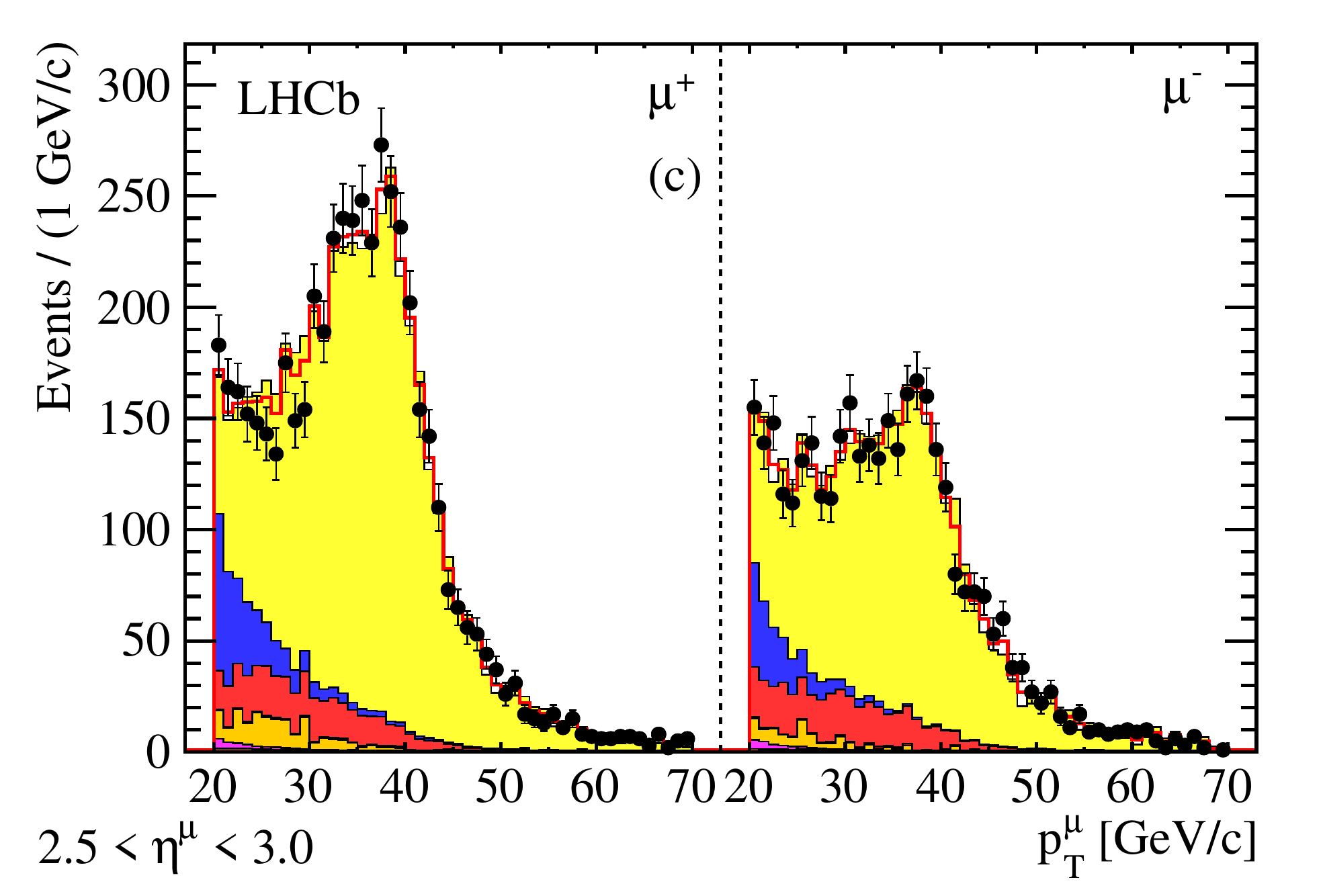}
\includegraphics[width=0.49\textwidth]{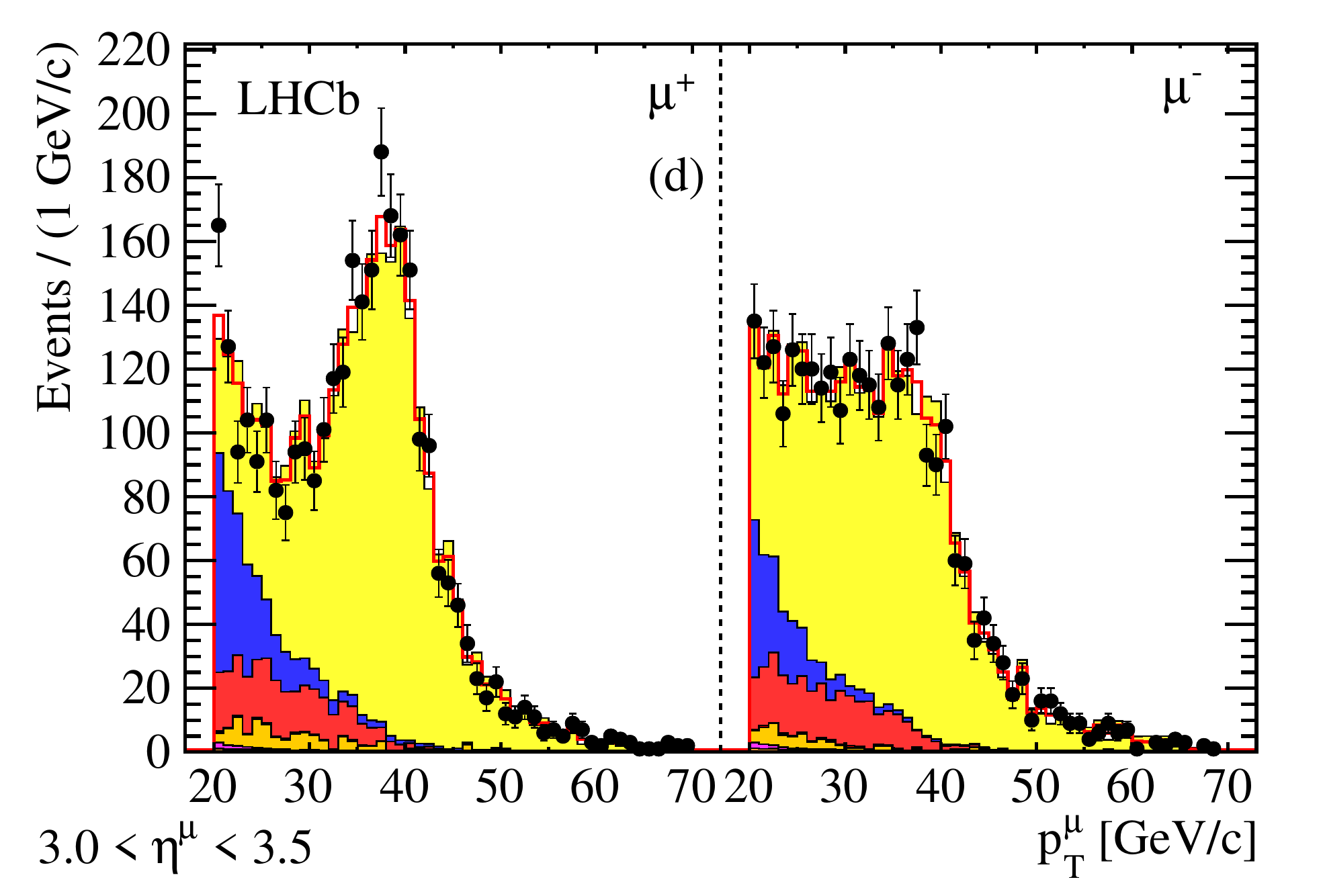}
\includegraphics[width=0.49\textwidth]{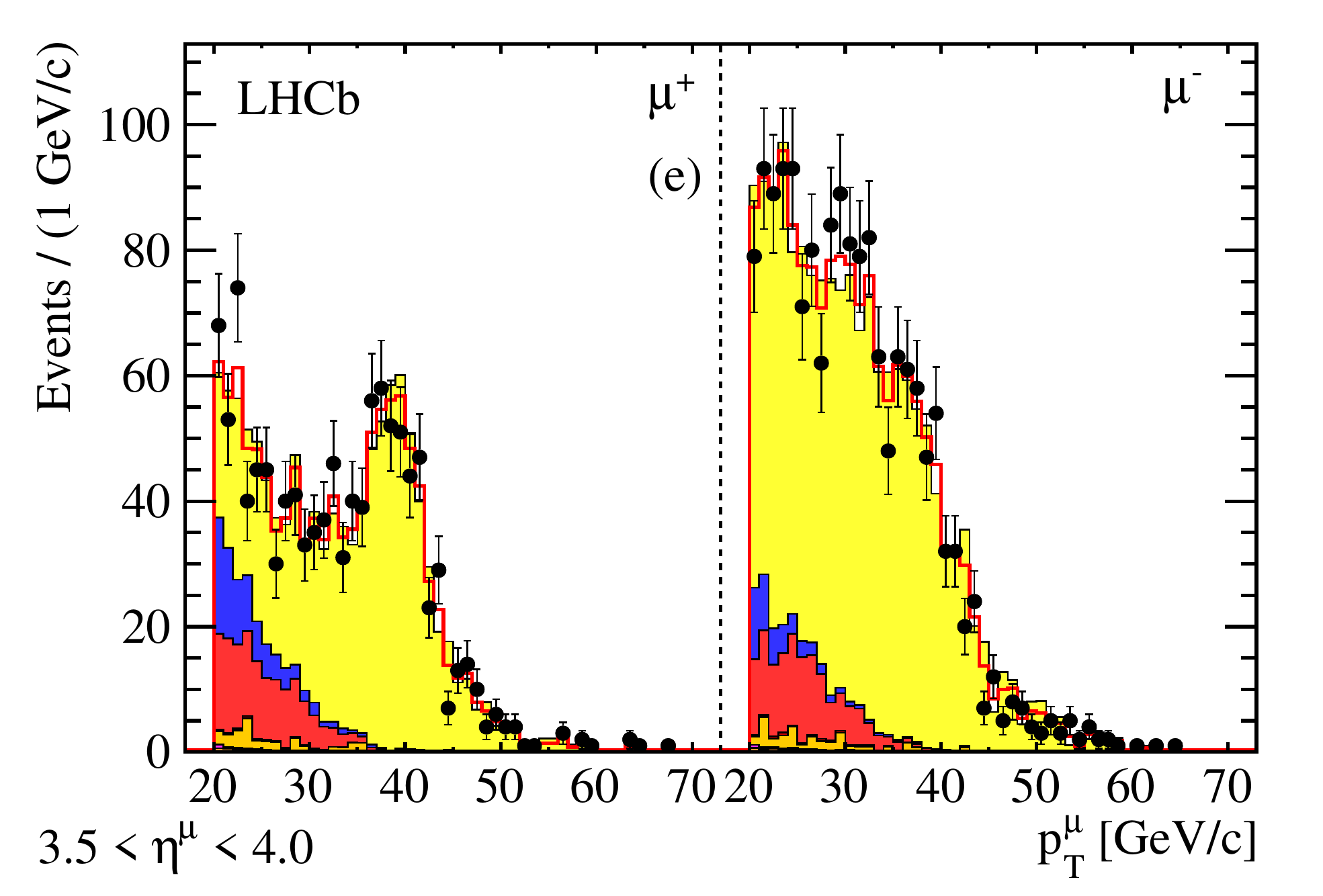}
\includegraphics[width=0.49\textwidth]{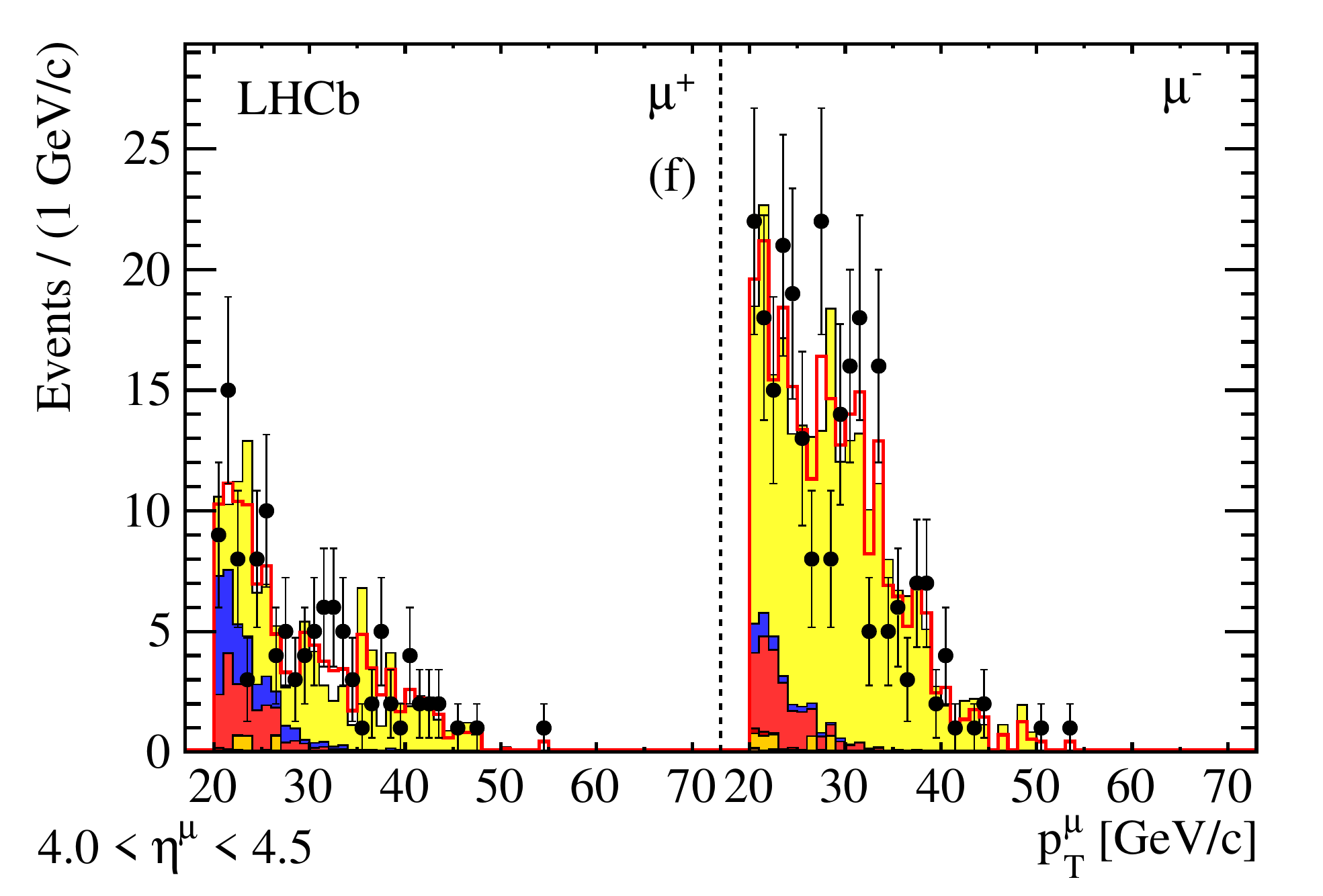}
\end{center}
\caption{
Distribution of muon $p_\mathrm{T}$ for positively (left panel) and negatively (right panel) 
charged muons in $W$ candidate events, 
for  the total fiducial cross-section (a). The plots (b) to (f) give the same information for the different $\etamu$ bins.
The data (points) are compared to the fitted contributions from $W^-$ and $W^+$ (light shaded). 
The background contributions are, from top to bottom in the legend:
decay-in-flight,  \zmumu, $\tau$ decays of \ww and $Z$, and heavy flavour decays.}
\label{fig:wfit}
\vspace*{1cm}
\end{figure}
The default fit has $15$ free parameters: five parameters for the normalisation of $W^+$
in each of the pseudorapidity bins, five parameters for $W^-$, and five parameters for the contribution 
coming from the decay-in-flight.  The normalisation of the other sources is fixed.
The result of the fit is shown in Fig.~\ref{fig:wfit}.
Integrated over both charges and $\ptmu$ it is found that 
  ($44.3\pm1.2$)\%
of the total sample is due to $W^+$,  ($34.9\pm1.1$)\% due to $W^-$,  ($8.5\pm0.8$)\% due to the decay-in-flight
contribution and the remainder due to the other backgrounds. The $\chi^2/$ndf of the fit is $1.002$.
The fit is repeated with the  \zmumu and \wmunu template corrected with {\sc Resbos} instead of NNLO, yielding
 ($43.6\pm1.2$)\% for  $W^+$ and ($34.4\pm1.1$)\% for $W^-$ with   $\chi^2/\mathrm{ndf}=0.983$.
The average of the two fits, which gives a purity $\rho^{W^+}=0.788\pm0.021$ for $W^+$ and $\rho^{W^-}=0.784\pm0.025$ for $W^-$, is taken for the final result; half of the difference is  taken as the systematic uncertainty.

 \section{Cross-section measurement}

\subsection{Cross-section definition}
Cross-sections are quoted in the kinematical range defined by the measurements.
The cross-sections are measured in bins of $\etamu$ for the $W$ and in bins of  $y^Z$ in case of the $Z$. 
The   cross-section  in a given bin of $y^Z$ ($\etamu$) is defined as
\begin{equation}
\sigmaz(y^Z)=\frac{\rho^Z f^Z_{\mathrm{FSR}}}{\mathcal{L} \mathcal{A}^Z }\sum\limits_{\eta_i^\mu,\eta_j^\mu}{\frac{N^Z(\eta_i^\mu,\eta_j^\mu)}{ \varepsilon^Z(\eta_i^\mu,\eta_j^\mu)}} , \mbox{~~~~~}
\sigmaw(\etamu)=\frac{\rho^W f^W_{\mathrm{FSR}}N^W }{\mathcal{L} \mathcal{A}^W \varepsilon^W },
\end{equation}
where $N^Z(\eta_i^\mu,\eta_j^\mu)$ is the number of $Z$ candidates in the respective $y^Z$ bin with the two muons in the bins $\eta_i^\mu$ and $\eta_j^\mu$
  being reconstructed with the efficiency $\varepsilon^Z(\eta_i^\mu,\eta_j^\mu)$.
Similarly, $N^W$ is the number of $W$ candidates with the muon in the $\etamu$ bin. 
The purity  of the sample ($\rho^{Z(W)}$), the acceptance ($\mathcal{A}^{Z(W)}$), the correction factor for final state radiation (FSR)  ($f^{Z(W)}_{\mathrm{FSR}}$) and the efficiency ($\varepsilon^W$) are determined per bin; $\mathcal{L}$ is the integrated luminosity.
The total cross-section is obtained by summing the contributions of the five  $y^Z$ or $\etamu$ bins.

\subsection{Signal efficiencies} 

The data are corrected for efficiency losses due to track reconstruction, muon identification, and
trigger requirements for both analyses.  There is an additional selection efficiency in the
$W$ analysis due to the requirements on the number of additional muons,
\IP, $E/pc$, $p_\mathrm{T}^\mathrm{cone}$, $E_\mathrm{T}^\mathrm{cone}$, and on TT hits.  
All efficiencies are determined from data.

The efficiencies for track reconstruction and muon identification are obtained 
using a tag-and-probe method  in the $Z$ sample. 
One of the muons in the $Z$ sample (tag) satisfies all the track criteria. 
The other muon (probe) is selected with looser criteria that depend on the efficiency to be measured. 
The invariant mass of the dimuon candidates, reconstructed from the tag and the probe muons, must lie in the window of $20$\gevcc around the nominal $Z$ mass.
The tracking efficiency, which accounts for track reconstruction and the track quality requirements, 
is studied using well reconstructed tracks in the muon stations which are linked to hits in TT.
The average track finding efficiency is about
$90$\% in the $Z$ sample and about $86$\% for the muon in $W$ events. The tracking efficiency for $W$ is lower due to the more restrictive cuts on the track quality.
The muon identification efficiency is determined with tracks without the muon identification requirement for the probe muon.
The average single muon efficiency  is above $99$\%. 
 Both the tracking and the muon identification efficiencies  agree  with simulation within errors.

The trigger efficiency contains two components, the first due to the efficiency of the single muon trigger and the other  due to the global event cuts (GEC).
The single muon trigger efficiency is  determined using the \zmumu sample.
 One muon is required to fire the single muon trigger. 
The trigger response of the 
other muon then defines the trigger efficiency. 
 The requirement on  the occupancy of the events depends on the multiplicity of 
the primary interactions.
It was checked with a sample which did not have the GEC applied, that 
no events are lost if there is only one primary vertex reconstructed. 
The GEC efficiency as a function of the number of primary 
vertices is determined by adding randomly triggered events to events with  $Z$ ($W$) candidates with one primary vertex; 
on average it amounts to $93$\%. 
The overall trigger efficiency is calculated for each event depending on the lepton pseudorapidity and the
primary vertex multiplicity. 
It is found to be about $88$\% for the $Z$ and  $75$\% for the $W$ sample.

The $W$ selection requires that there are no other muons with $\ptmu>2$\gevc,
\mbox{$p_\mathrm{T}^\mathrm{cone}<2$\gevc,} \mbox{$E_\mathrm{T}^\mathrm{cone}<2$\gev}, \mbox{\IP$<40$\mum}, and $E/pc<0.04$.
The selection efficiency is determined from the fraction
  of \pseudow events that pass these requirements.
A similar method is used to evaluate the efficiency for the requirement of TT hits associated to the muon track of the $W$ candidate.
Simulation studies show that with the exception of the $E/pc$ distribution, the \pseudow
data provide a consistent description of \wmunu simulation, as shown
in Figs.~\ref{fig:conehf} and \ref{fig:life}.  
However, the harder muon $p_\mathrm{T}$ spectrum in \pseudow data leads to slightly lower values
of $E/pc$ than for muons produced in $W$ decays.  The simulation is used to determine this
difference, which is only significant for $\etamu$ between $2.0$ and $2.5$, where the efficiency
for $W$ events is estimated to be $2.1$\% lower than for \pseudow data.
The selection efficiency is about $67$\% for  $2.5<\etamu<4.0$
and drops to about $53$\% and $33$\%  for the two bins at the edge of the acceptance 
with $2.0<\etamu<2.5$ and $4.0<\etamu<4.5$, respectively. 

All the efficiencies have been checked for possible dependences on $p_\mathrm{T}^\mu$, the azimuthal angle of the muon, magnet polarity, and  $\eta^\mu$. Only the latter exhibits a significant dependence, which is taken into account. 
Since any charge bias of the efficiencies would directly influence the measurement of the lepton charge asymmetry, it was checked 
there is no significant charge dependence within the uncertainties of the efficiencies.
The efficiency corrections are applied as a function of the pseudorapidity of the muons  except the GEC.

The efficiencies are uncorrelated between pseudorapidity bins  but correlated for $W^+$, $W^-$ and $Z$. 
These correlations are taken into account for the measurement of the lepton charge asymmetry and the cross-section ratios.

\subsection{Acceptance}

The selection criteria for the $W$ and $Z$  define the fiducial region of the measurement.
Simulated events are used to determine the acceptance $\mathcal{A}$, 
 defined as $\mathcal{A}={N_{\mathrm{rec}}}/{N_{\mathrm{gen}}}$.
Here, $N_{\mathrm{rec}}$ is the number of reconstructed events satisfying the cuts on the  pseudorapidity 
and the minimal momenta of the reconstructed muons, as well as on  the dimuon mass in the case of the $Z$ analysis.
Similarly,  $N_{\mathrm{gen}}$ is the number of generated events with the cuts applied on the generated muons. 
The acceptance is estimated with $W$-MC and $Z$-MC. It is found to be consistent with unity for the $Z$  and above $0.99$ for the $W$ analysis. For the latter, the acceptance corrects for the small loss of events with  $\ptmu>70$\gevc.

\subsection{Luminosity}
The absolute luminosity scale was measured at specific periods during the data taking, using both Van der Meer scans~\cite{vandermeer} where colliding beams
are moved transversely across each other to determine the beam profile, and a beam-gas imaging method~\cite{beamgas,lhcblumi}.
For the latter, reconstructed beam-gas interaction vertices near the 
beam crossing point determine the beam profile.
Both methods give similar results and are estimated to have a precision of order $3.5$\%.
The knowledge of the absolute luminosity scale is used to calibrate the number of tracks in the VELO, which is found to be stable throughout the data-taking period and can therefore be used to monitor the instantaneous luminosity of the entire data sample.  
The dataset for this analysis corresponds to 
an integrated luminosity of $37.1\pm1.3$\invpb.

\subsection{Corrections to the data}\label{sec:corrections}

The measured cross-sections are corrected to Born level in quantum electrodynamics (QED) 
in order to provide a consistent comparison with NLO and NNLO QCD predictions, 
which do not include the effects of FSR. 
Corrections have been  estimated using {\sc Photos}~\cite{photos} interfaced to {\sc Pythia}.
The {\sc Pythia} $p_\mathrm{T}$ spectrum of the electroweak boson has been reweighted  to the NNLO spectrum as determined with {\sc Dynnlo}~\cite{dynnlo}.
The correction is taken as the number of events within the fiducial cuts of the measurements after FSR divided by the number of events generated within the fiducial cuts.

{\sc Pythia} simulation is used to study    bin-to-bin migrations for  $\etamu$ and $y^Z$.
 No significant net migration is observed and 
no correction is applied. 

\subsection{Systematic uncertainties}\label{sec:syst}
\begin{table}[!tb]
\caption{Contributions to the systematic uncertainty for the total $Z$ and $W$ cross-sections.  The different contributions are discussed in Sect.~\ref{sec:syst}} 
\begin{center}
\resizebox{0.98\textwidth}{!}{
\begin{tabular}{l|c|c|c}
Source               & $\Delta \sigmaz$ (\%)& $\Delta \sigmawp$ (\%)& $\Delta \sigmawm$ (\%)\\ \hline
Signal purity               & $\pm0.1$ & $\pm1.2 $  &  $\pm0.9$ \\
Template shape (fit) & -- & $\pm0.9$ & $\pm1.0$  \\
Efficiency (trigger, tracking, muon id) & $\pm4.3$& $\pm2.2$ & $\pm2.0$ \\
Additional selection            & -- & $\pm1.8 $& $\pm1.7 $\\
FSR correction       & $\pm0.02$& $\pm0.01$ & $\pm0.02$  \\ \hline
Total            & $\pm4.3$& $\pm3.2$ & $\pm2.9$  \\ \hline
Luminosity          & $\pm3.5$ & $\pm3.5$ & $\pm3.5$ \\ 
\end{tabular}}
\end{center}
\label{tab:sys}
\end{table}
Aside from the uncertainty on the luminosity measurement, the main sources of experimental 
uncertainties come from the efficiency determinations and the 
  background estimate in the $W$ analysis.
The following sources have been considered:

\begin{enumerate}
\item The relative uncertainties of the tracking, muon identification, trigger and GEC efficiencies 
are added in quadrature. 
They lead to a systematic uncertainty for the total cross-sections of $4.3$\% ($2.2$\%, $2.0$\%) for the $Z$ ($W^+$, $W^-$).

\item The statistical uncertainty on the efficiency of the additional selection cuts for the $W$ analysis
translates into a $1.8$\% ($1.7$\%) systematic uncertainty on the total   $W^+$ ($W^-$) cross-section.

\item The uncertainty of the background contribution 
for the $Z$ analysis is small; the  uncertainty  in the determination
of the sample purity leads to a  $0.1$\% uncertainty on the total cross-section.

\item Both the shape and normalisations of the templates used in the $W$
 fit are considered as an additional source of uncertainty. To determine this   
 systematic uncertainty each of the following sources 
is varied in turn, the
data are refitted to determine the fraction of $W^+$ and $W^-$ events, and the deviations from the original signal yield are combined in quadrature.
The following variations are made:
\begin{itemize}
\item the difference of the two fits using different  $W$ and $Z$ templates (see Sect.~\ref{sec:yield}) 
leads to a variation on the $W$ fractions of $0.8$\%;
\item the normalisation of the $Z$ component was changed by the statistical uncertainty
with which it was determined, leading to a variation in the $W$ fractions of $0.3$\%;
\item the normalisation of the $W\rightarrow \tau\nu$ template was changed by the statistical
uncertainty with which it was determined, leading to a negligible change in the $W$ fractions,
since this template shape is very similar to the decay-in-flight template which is allowed to vary
in the fit;
\item the heavy flavour template has also been changed by the statistical uncertainty with
which it was determined leading to a negligible change in the $W$ fractions;
\item instead of leaving the relative normalisation of the decays-in-flight template
  between pseudorapidity bins to be free in the fit,
 this is fixed to the values observed in randomly triggered events, and the  
 full fit performed with a single free parameter for the background; the
 $W$ fractions
change by $0.2$\%;
\item the shape of the decay-in-flight template has been changed using
different values for $\alpha$~(see Eq.~\ref{equ:fit}) to describe the decay probability, corresponding
to different regions in which the hadron must have decayed;\footnote{Three different decay regions have been considered: from the interaction point, from the VELO and from the TT stations up to the electromagnetic calorimeter.}
 no difference in the $W$ fractions is observed.

\end{itemize}
\item The uncertainty on the FSR correction is evaluated for each bin as the maximum of the statistical uncertainty of the correction factor and 
 the difference between the weighted and unweighted FSR correction factor. 

\end{enumerate}
The sources of systematic uncertainties are summarised in Table~\ref{tab:sys}, together with
the size of the resultant uncertainty on the $W$ and $Z$ total cross-sections.
The total systematic uncertainty is the sum of all contributions added in quadrature.

 \section{Results}

The  inclusive cross-sections for \zmumu and  \wmunu production 
for muons with $\ptmu>20$\gevc in the pseudorapidity region $2.0<\etamu<4.5$ and, in the case of $Z$, the invariant mass range $60<M_{\mu\mu}<120$\gevcc are measured to be 

\begin{center}\begin{tabular}{lcl}
$\sigmaz$& =& \zxsec\\
$\sigmawp$&=& \wplusxsec \\
$\sigmawm$ &=& $656\pm8\pm19\pm 23$\pb\,\!\!,\\
\end{tabular}\end{center}
\begin{figure}[!tbp]
  \begin{center} 
    \includegraphics[width=0.88\textwidth]{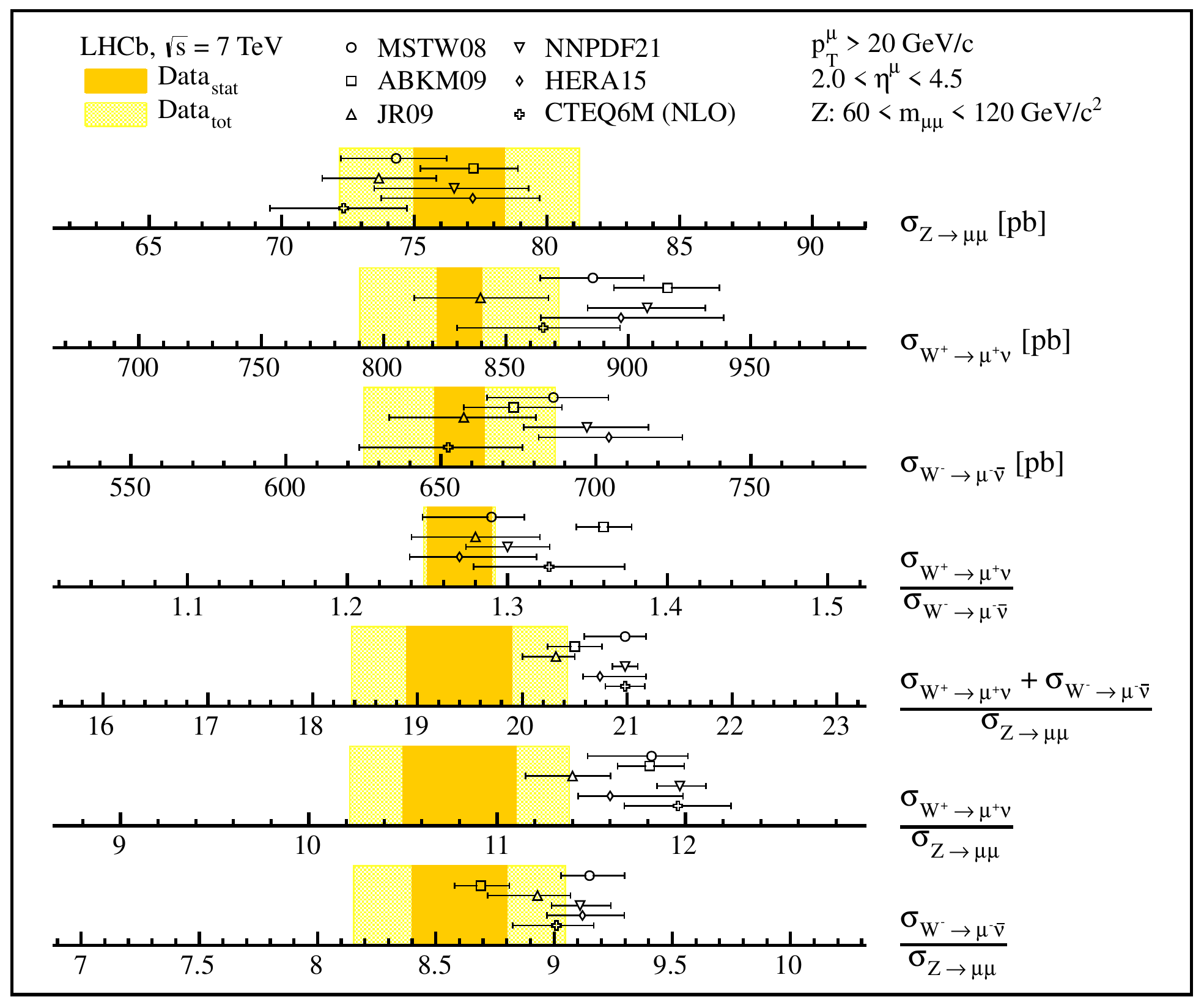}
    \caption{Measurements of the $Z$, $W^+$ and $W^-$ cross-section  and ratios, data are shown as bands which  the statistical (dark shaded/orange) and total (light hatched/yellow) errors. The measurements are compared to NNLO and NLO predictions with different PDF sets for the proton, shown as points with error bars. 
The PDF uncertainty, evaluated at the $68$\% confidence level, and the theoretical uncertainties are added in quadrature to obtain the uncertainties of the predictions.}
    \label{fig:cfall}
  \end{center}
\end{figure}
where the first uncertainty is statistical, the second  systematic  and the third is due to the luminosity.
All the measurements are dominated by  the luminosity and the systematic uncertainty. 
The latter is dominated by the limited number of events for the background templates and in the determination of the efficiencies.

The ratios $R_W=\sigmawp/\sigmawm$ 
and  $R_{WZ}=(\sigmawp+\sigmawm)/\sigmaz$
 are measured to be
\begin{center}\begin{tabular}{lcl}
$R_{W}$ &=& \wratio\\
$R_{WZ}$& =& \wzratio\,.\\
\end{tabular}\end{center}
Here, the uncertainty from the luminosity  completely cancels. 
The systematic uncertainty on  the trigger, muon identification, tracking and selection efficiencies, as well as the uncertainty on the purity are assumed to be fully correlated between $W^+$ and $W^-$. No correlation is assumed between the $\etamu$ bins, except for the purity.
The uncertainty on the $Z$ cross-section from the reconstruction
 efficiency is correlated between  boson rapidity bins.
The correlation of the uncertainty on the efficiencies between $W$ and $Z$ are estimated with MC simulation to be  $90$\%. The full correlation matrix is given in   the Appendix (Table~\ref{tab:corr}).
The ratio of the $W$ to $Z$  cross-section is measured, for each charge separately, to  be
\begin{center}\begin{tabular}{lcl}
$\sigmawp/\sigmaz$& =& $10.8\pm0.3\pm 0.5$\\
$\sigmawm/\sigmaz$&=& $8.5\pm0.2\pm 0.4$\,.\\
\end{tabular}\end{center}

A summary of the measurements of the inclusive cross-sections $\sigmawp$, 
$\sigmawm$ and $\sigmaz$, 
and the ratios  is shown in Fig.~\ref{fig:cfall}. 
The measurements are shown as a band which represents
 the total and statistical uncertainties.


\begin{figure}[!htbp]
  \begin{center} 
    \includegraphics[width=0.72\textwidth]{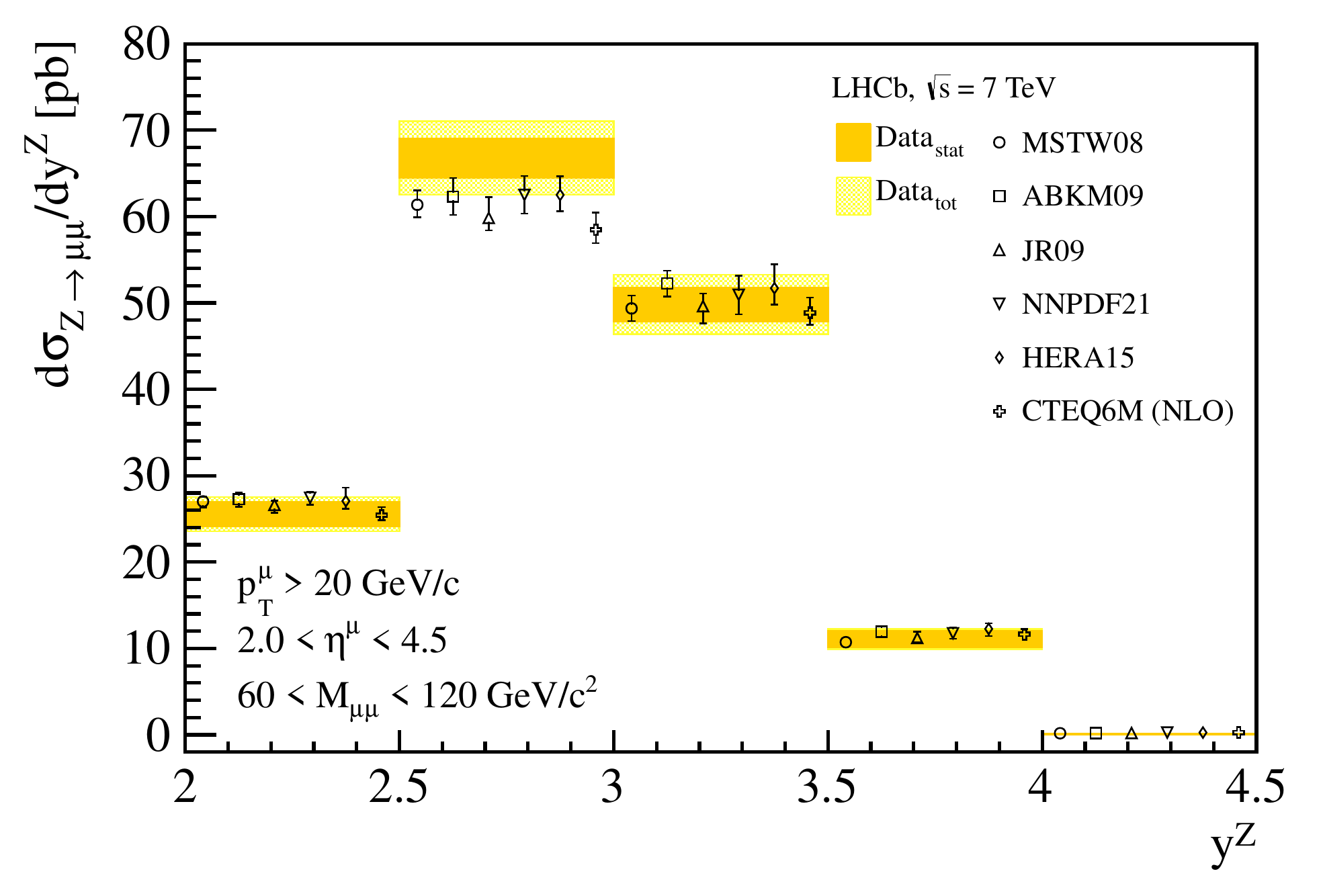}
    \caption{Differential cross-section for \zmumu as a function of $y^Z$. The  dark shaded (orange) bands correspond to the statistical uncertainties, the light hatched (yellow) band to the statistical and systematic uncertainties added in quadrature.
Superimposed are NNLO (NLO) predictions with different parametrisations for the PDF as points with error bars; they are displaced horizontally for presentation.  }
    \label{fig:z_rap}
  \end{center}
  \begin{center} 
    \includegraphics[width=0.72\textwidth]{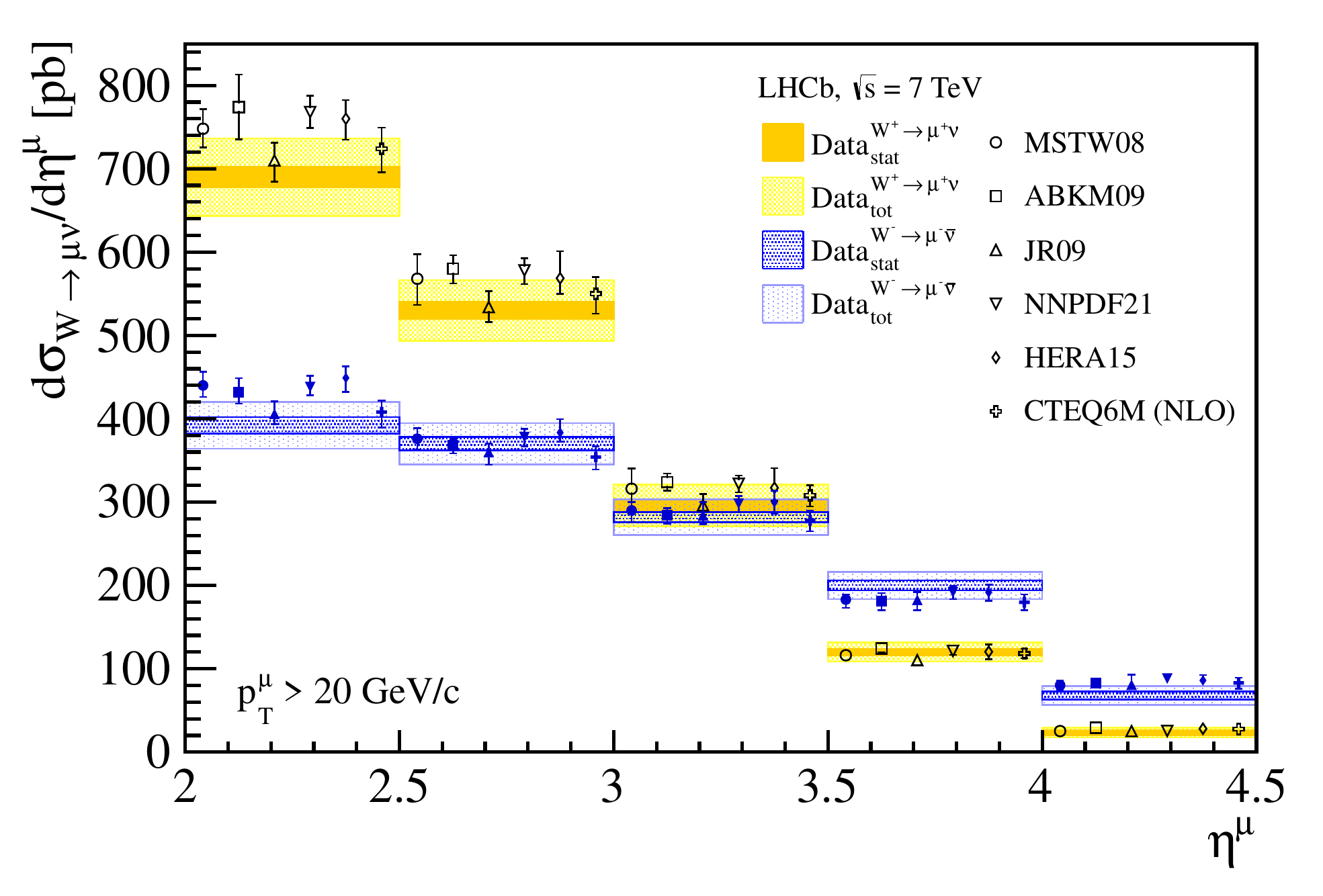}
    \caption{Differential $W$ cross-section in bins of muon pseudorapidity.
The  dark shaded (orange) bands correspond to the statistical uncertainties, the light hatched (yellow) band to the statistical and systematic uncertainties added in quadrature.
Superimposed are NNLO (NLO) predictions as described in Fig~\ref{fig:z_rap}.  }
    \label{fig:wxsec}
  \end{center}
\end{figure}
\begin{figure}[!htbp]
  \begin{center} 
    \includegraphics[width=0.72\textwidth]{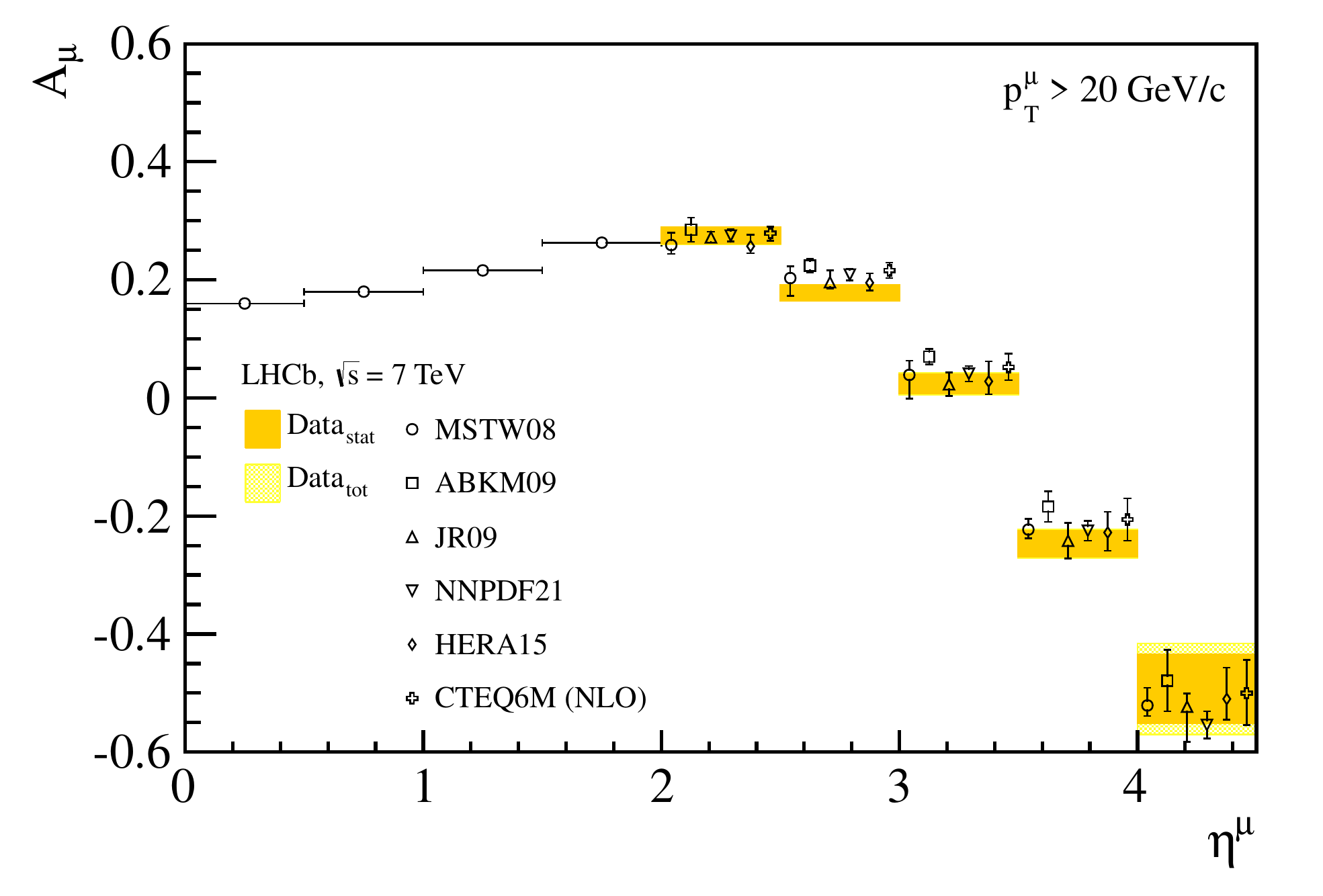}
    \caption{Lepton charge asymmetry $A_\mu=(\sigmawp-\sigmawm)/(\sigmawp+\sigmawm)$ in bins of muon pseudorapidity.
The  dark shaded (orange) bands correspond to the statistical uncertainties, the light hatched (yellow) band to the statistical and systematic uncertainties added in quadrature.
Superimposed are NNLO (NLO) predictions as described in Fig~\ref{fig:z_rap}. 
The \MSTW values for $\etamu<2$ represent the central value of the prediction.}
    \label{fig:cfdiff}
  \end{center}
\end{figure}
The results are compared to 
theoretical predictions calculated at NNLO with the program {\DYNNLO~\cite{dynnlo} for the NNLO PDF sets of \MSTW~\cite{mstw08}, \ABKM~\cite{abkm09}, \JR~\cite{jr09}, \HERA~\cite{h1zeus} and   \NNPDF~\cite{nnpdf} and at NLO for the NLO PDF set \CTEQM~\cite{cteq}.\footnote{\DYNNLO sets $\alpha_s$ to the value of $\alpha_s$ at the mass of the $Z$ boson as given by the different PDF sets.}
The scale uncertainties are estimated by varying the renormalisation and factorisation scales by factors of two around the nominal value, which is set to the boson mass. The uncertainties for each set correspond to the PDF uncertainties at 68\% and the  scale uncertainties added in quadrature.\footnote{The uncertainties for the PDF set from  \CTEQM which is given at $90$\% CL are divided by $1.645$. }

While the $W^-$ and $Z$ cross-sections are well described by all predictions, the $W^+$ cross-section is slightly
overestimated by the \ABKM  and  \NNPDF PDF sets. 
The ratio of the $W^-$ to  $Z$ cross-sections agrees  reasonably well with the predictions, but the   $W^+$ to the $Z$ ratio is overestimated by most of the predictions.  
The systematic uncertainties for the $R_W$  almost cancel and also the theoretical uncertainties are much reduced.
The  $R_W$ measurement tests the Standard Model predictions 
 with a precision of $1.7\%$ which is comparable to the uncertainty of the theoretical prediction.
 The \ABKM prediction overestimates this ratio while all the other predictions agree with the measurement.
\begin{figure}[!htbp]
  \begin{center} 
    \includegraphics[width=0.72\textwidth]{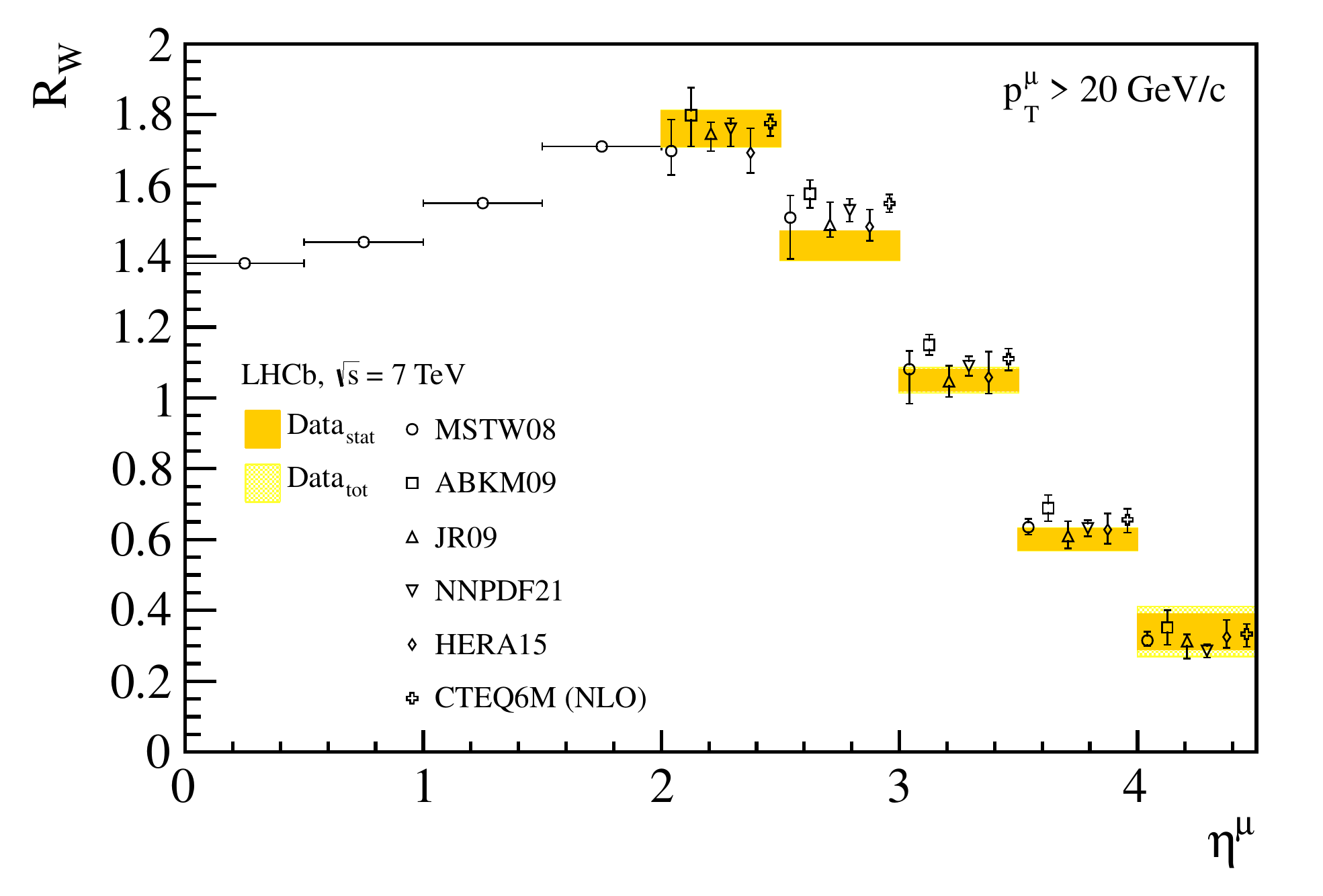}
    \caption{$R_W=\sigmawp/\sigmawm$  in bins of muon pseudorapidity.
The  dark shaded (orange) bands correspond to the statistical uncertainties, the light hatched (yellow) band to the statistical and systematic uncertainties added in quadrature.
Superimposed are NNLO (NLO) predictions with different parametrisations as described in Fig~\ref{fig:z_rap}.  
The  \MSTW values for $\etamu<2$ represent the central value of the prediction.}
    \label{fig:wratio}
  \end{center}
\end{figure}
Differential distributions are measured in five bins in $y^Z$ for the \textit{Z} and of $\etamu$ for the \textit{W}.
Figure~\ref{fig:z_rap} shows the differential cross-section as a function of the rapidity  
of the $Z$ boson together with NNLO (NLO) predictions with different  parametrisation for the PDFs of the proton.
 The predictions agree with the measurements within uncertainties though all the predictions are lower than the measured cross-section for $2.5<\etamu<3.0$. 
The differential cross-sections are listed in Table~\ref{tab:z_rap} in the Appendix.

The differential distribution of the $W^+$ and $W^-$ cross-section,  the lepton charge asymmetry $A_\mu$ and the ratio $R_W$
 as a function of the muon pseudorapidity are shown in Figs.~\ref{fig:wxsec},~\ref{fig:cfdiff} and ~\ref{fig:wratio} and listed in Tables~\ref{tab:w_rap} to \ref{tab:wratio} as a function of $\ptmu$. 
 The measurement of the charge asymmetry and the $W$ ratio provides important additional information on the PDFs particularly on the 
 valence quark distributions~\cite{halzen}. 

 Since the inclusive  cross-section for $W^+$ is larger than for $W^-$, due to the excess of $u$ over $d$ quarks in the proton, the overall asymmetry is positive. 
 The asymmetry and the $W$ cross-sections strongly vary  as a function of the pseudorapidity of the charged lepton, and $A_\mu$ even changes sign, owing to differing helicity dependence of the lepton couplings to the 
boson. This behaviour is reflected in the differential $W$ cross-sections, where at  large muon pseudorapidities the $W^-$ cross-section is higher than the $W^+$  cross-section, as a consequence of  the $V-A$ structure of the $W$ to lepton coupling.
The cross-section  and the asymmetry measurements  are compared to the NNLO (NLO) predictions with different parameterisation for the PDFs.
 The \ABKM prediction overestimates the measured asymmetry in 
three of the five bins.
The other predictions describe the measurement within uncertainties.
\begin{figure}[!thtp]
  \begin{center} 
    \includegraphics[width=0.72\textwidth]{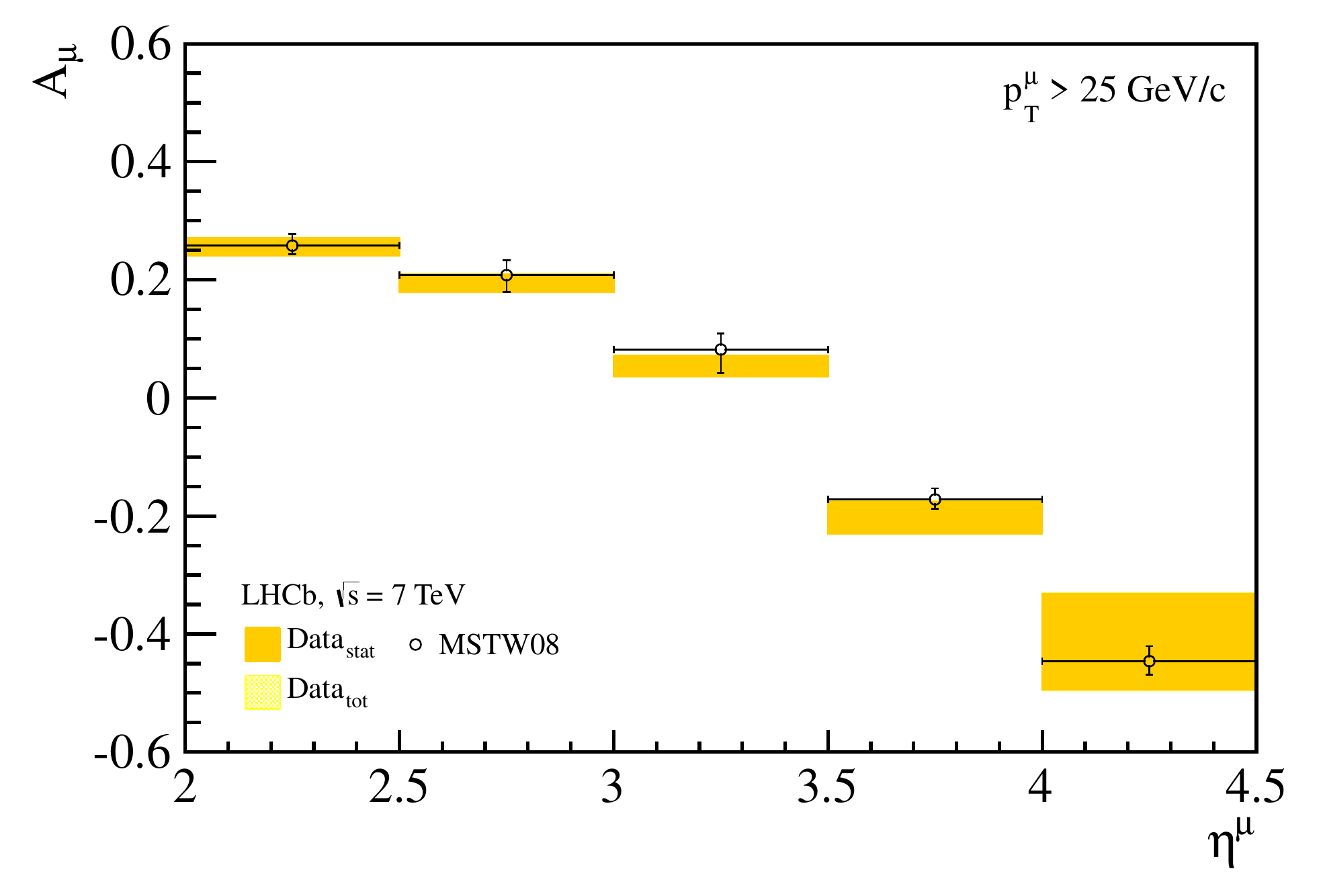}
    \includegraphics[width=0.72\textwidth]{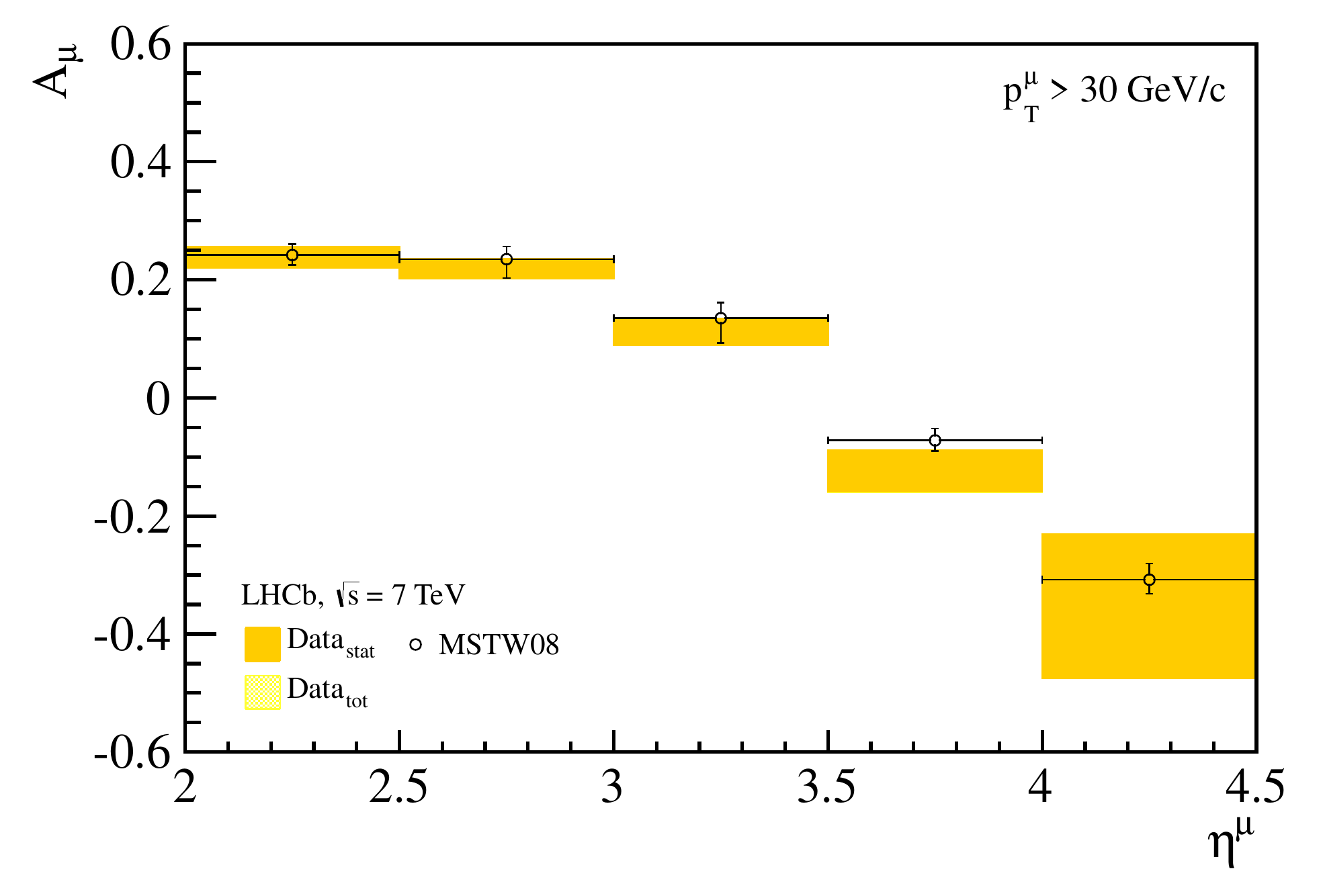}
    \caption{Lepton charge asymmetry $A_\mu=(\sigmawp-\sigmawm)/(\sigmawp+\sigmawm)$ for muons with $\ptmu>$25 (top) and 30\gevc (bottom), respectively in bins of muon pseudorapidity.
The  dark shaded (orange) bands correspond to the statistical uncertainties, the light hatched (yellow) band to the statistical and systematic uncertainties added in quadrature.
 The statistical uncertainty  is undistinguishable from the total uncertainty. 
Superimposed are the  NNLO predictions with the \MSTW parametrisation for the PDF.\vspace*{1.5cm}}
    \label{fig:asy2}
  \end{center}
\end{figure}

The asymmetry is also measured for two higher $\ptmu$ thresholds for the muons, 
at 25  and 30\gevc.
The result is shown in  Fig.~\ref{fig:asy2} and listed in Table~\ref{tab:asy2}. The NNLO prediction with \MSTW parametrisation for the PDF also describes the measured
asymmetry with the higher cuts on the transverse momentum of the muon. 

 
\section{Conclusions}
Measurements  of  inclusive  $W$ and $Z$ boson production in $pp$ collisions 
at $\sqrt{s}=7$\tev with final states containing muons have been performed using $37$\invpb of 
data collected with the LHCb detector.
The inclusive cross-sections have been measured separately for $W^+$ and $W^-$ production as well as the ratios  $\sigmawp/\sigmawm$ 
and $(\sigmawp+\sigmawm)/\sigmaz$ and the lepton charge asymmetry $(\sigmawp-\sigmawm)/(\sigmawp+\sigmawm)$.
The results have been compared to five next-to-next-to-leading order QCD predictions with different sets for the parton density 
functions of the proton and to one calculation at next-to-leading order. 
There is general agreement with the predictions, though  some of the PDF sets overestimate the ratios of the cross-sections.
The ratio $\sigmawp/\sigmawm=$\wratio\, is measured precisely and allows  the Standard Model prediction to be tested with an accuracy of about $1.7$\%, comparable to the uncertainty on the theory prediction.
These represent the first measurements of the $W$ and $Z$ production cross-sections and ratios in
the forward region at the LHC, and will provide valuable input to the knowledge of the parton density
functions of the proton.
The uncertainty on the cross-section measurements is dominated by systematic uncertainties. Since most of these are statistical in nature, the accuracy on the
measurement with further data is expected to significantly improve.

\section*{Acknowledgements}

\noindent We express our gratitude to our colleagues in the CERN accelerator
departments for the excellent performance of the LHC. We thank the
technical and administrative staff at CERN and at the LHCb institutes,
and acknowledge support from the National Agencies: CAPES, CNPq,
FAPERJ and FINEP (Brazil); CERN; NSFC (China); CNRS/IN2P3 (France);
BMBF, DFG, HGF and MPG (Germany); SFI (Ireland); INFN (Italy); FOM and
NWO (The Netherlands); SCSR (Poland); ANCS (Romania); MinES of Russia and
Rosatom (Russia); MICINN, XuntaGal and GENCAT (Spain); SNSF and SER
(Switzerland); NAS Ukraine (Ukraine); STFC (United Kingdom); NSF
(USA). We also acknowledge the support received from the ERC under FP7
and the Region Auvergne.

 \clearpage
\bibliographystyle{LHCb}
\bibliography{main}

 \clearpage

{\noindent\bf\Large Appendix}

\appendix

\section{Tables of results}
\begin{table}[h]
\caption{Correlation coefficients between  $W^+$, $W^-$ and \textit{Z} in the five bins considered. The luminosity uncertainty is not included.}
\begin{center}

{\scriptsize
\renewcommand{\arraystretch}{2.5}
\begin{tabular}{l|ccc|ccc|ccc|ccc|ccc|c}
& \multicolumn{3}{|c|}{$2 < \etamu \, (y^Z) < 2.5$} & \multicolumn{3}{|c|}{$2.5 < \etamu \, (y^Z) < 3$} & \multicolumn{3}{|c|}{$3 < \etamu \, (y^Z) < 3.5$} & \multicolumn{3}{|c|}{$3.5 < \etamu \, (y^Z) <
4$} & \multicolumn{3}{|c|}{$4 < \etamu \, (y^Z) < 4.5$} \\ \hline

\Wp & 1 &&&&&&&&&&&&&&& \multirow{3}{*}{\begin{sideways}$2 < \etamu \, (y^Z) < 2.5$\end{sideways}} \\
\Wm & 0.87 & 1 &&&&&&&&&&&&&& \\
$Z$ & 0.36 & 0.34 & 1 &&&&&&&&&&&&& \\ \hline

\Wp & 0.02 & 0.02 & 0.35 & 1 &&&&&&&&&&&& \multirow{3}{*}{\begin{sideways} $2.5 < \etamu \, (y^Z) < 3$\end{sideways}}\\
\Wm & 0.02 & 0.02 & 0.35 & 0.90 & 1 &&&&&&&&&&& \\
$Z$ & 0.47 & 0.44 & 0.45 & 0.45 & 0.45 & 1 &&&&&&&&&& \\ \hline

\Wp & 0.02 & 0.03 & 0.24 & 0.02 & 0.02 & 0.31 & 1 &&&&&&&&& \multirow{3}{*}{\begin{sideways} $3 < \etamu \, (y^Z) < 3.5$\end{sideways}} \\
\Wm & 0.02 & 0.02 & 0.29 & 0.02 & 0.02 & 0.37 & 0.89 & 1 &&&&&&&& \\
$Z$ & 0.46 & 0.43 & 0.44 & 0.45 & 0.44 & 0.58 & 0.31 & 0.37 & 1 &&&&&&& \\ \hline

\Wp & 0.04 & 0.05 & 0.35 & 0.04 & 0.04 & 0.45 & 0.05 & 0.04 & 0.44 & 1 &&&&&& \multirow{3}{*}{\begin{sideways} $3.5 < \etamu \, (y^Z) < 4$\end{sideways}} \\
\Wm & 0.02 & 0.02 & 0.40 & 0.02 & 0.01 & 0.52 & 0.02 & 0.02 & 0.51 & 0.80 & 1 &&&&& \\
$Z$ & 0.32 & 0.29 & 0.30 & 0.30 & 0.30 & 0.39 & 0.21 & 0.25 & 0.39 & 0.30 & 0.35 & 1 &&&& \\ \hline

\Wp & 0.07 & 0.09 & 0.19 & 0.07 & 0.07 & 0.24 & 0.09 & 0.07 & 0.24 & 0.15 & 0.06 & 0.16 & 1 &&& \multirow{3}{*}{\begin{sideways} $4 < \etamu \, (y^Z) < 4.5$\end{sideways}}\\
\Wm & 0.01 & 0.01 & 0.28 & 0.01 & 0.01 & 0.37 & 0.01 & 0.01 & 0.36 & 0.02 & 0.01 & 0.24 & 0.57 & 1 && \\
$Z$ & 0.03 & 0.03 & 0.03 & 0.03 & 0.03 & 0.04 & 0.02 & 0.03 & 0.04 & 0.03 & 0.04 & 0.03 & 0.02 & 0.03 & 1 & \\ \hline

& \Wp & \Wm & $Z$ & \Wp & \Wm & $Z$ & \Wp & \Wm & $Z$ & \Wp & \Wm & $Z$ & \Wp & \Wm & $Z$ & \\
\end{tabular}}
\end{center}
\label{tab:corr}
\end{table}

\begin{table}[!htbp]
\caption{Differential $Z\rightarrow \mu \mu$
cross-section, $d\sigmaz/dy^Z$,  in bins of boson rapidity. The first cross-section uncertainty is statistical, the second 
 systematic, and the third due to the uncertainty on the luminosity
 determination.
The correction factor $f^Z_{\mathrm{FSR}}$ which is used to correct for FSR is listed separately. 
}
\begin{center}
\begin{tabular}{c|clll|c}
$y^Z$  &\multicolumn{4}{|c|}{  $d\sigmaz/dy^Z$ [pb]} & $f^Z_{\mathrm{FSR}}$ \\ \hline
$2.0-2.5$ &  $25.5 $&$ \pm  1.4 $&$\pm 1.0 $&$ \pm   0.9$  &$1.020  \pm 0.001$  \\
$2.5-3.0$ &  $66.8 $&$ \pm  2.3 $&$\pm 2.7 $&$ \pm  2.3$   &$1.018 \pm  0.001$   \\
$3.0-3.5$ &  $49.8 $&$ \pm 2.0  $&$\pm 2.2 $&$ \pm 1.7$    &$1.018  \pm  0.001$  \\
$3.5-4.0$ &  $11.1  $&$ \pm 0.9  $&$\pm 0.6  $&$ \pm 0.4$    & $1.024 \pm  0.001$  \\
$4.0-4.5$ &  $0.074 $&$ \pm 0.074 $&$\pm 0.004 $&$ \pm 0.002$     &$1.027 \pm  0.027$  \\
\end{tabular}
\end{center}
\vspace{1cm}
\label{tab:z_rap}
\caption{Differential $W\rightarrow \mu \nu$ 
cross-section, $d\sigmaw/\etamu$,  in bins of lepton pseudorapidity. 
The first cross-section uncertainty is statistical, the second 
 systematic, and the third due to the uncertainty on the luminosity
 determination.
The correction factor $f^W_{\mathrm{FSR}}$ which is used to correct for FSR is listed separately. 
}
\begin{center}
\begin{tabular}{c|c|clll|c}

& $\etamu$ & \multicolumn{4}{|c|}{$d\sigmaw/\etamu$ [pb]} & $f^W_{\mathrm{FSR}}$ \\ \hline
$W^+$ &$2.0 - 2.5$ &  $691  $&$ \pm  12  $&$ \pm  37   $&$ \pm   24$ &$1.0146 \pm  0.0004$\\
&$2.5 - 3.0$       &  $530  $&$ \pm  9   $&$ \pm  30   $&$ \pm   19 $  &$1.0086 \pm  0.0002$\\
&$3.0 - 3.5$       &  $296  $&$ \pm  7   $&$ \pm  23   $ &$ \pm   10 $   &$1.0107 \pm  0.0006$\\
&$3.5 - 4.0$       &  $121   $&$ \pm  5   $&$ \pm  19   $&$ \pm   4 $  &$1.0097 \pm  0.0005$\\ 
&$4.0 - 4.5$       &  $23.1 $&$ \pm  3.2   $&$ \pm  4.9   $ &$ \pm  0.8  $ &$1.0009 \pm  0.0009$\\ 
\hline 
$W^-$ &$2.0 - 2.5$ &  $393  $&$ \pm  9  $&$ \pm  22  $&$ \pm  13$ &$1.0147 \pm  0.0008 $\\
&$2.5 - 3.0$       &  $370  $&$ \pm  8  $&$ \pm  20  $&$ \pm  13$  &$1.0163 \pm  0.0004 $\\
&$3.0 - 3.5$       &  $282  $&$ \pm  7  $&$ \pm  18  $&$ \pm  10$ &$1.0147 \pm  0.0004 $\\
&$3.5 - 4.0$       &  $200   $&$ \pm  6  $&$ \pm  14  $&$ \pm  7$ &$1.0173 \pm  0.0008 $\\
&$4.0 - 4.5$       &  $68   $&$ \pm  5  $&$ \pm  10  $&$ \pm  2$ &$1.0194 \pm  0.0009 $\\ 
\end{tabular}
\end{center}
\label{tab:w_rap}
\end{table}

\begin{table}[!tbp]
\caption{Lepton charge asymmetry, $A_\mu$, in bins of muon pseudorapidity
 for a $\ptmu$ threshold  at 20, 25 and 30 GeV/c. The first uncertainty is 
 statistical and the second systematic. The effect of FSR is at the level of 
 $10^{-4}$ and is not listed.
}
\begin{center}\resizebox{0.98\textwidth}{!}{
\begin{tabular}{c|cll|cll|cll}
$\eta^{\mu}$ & \multicolumn{3}{|c|}{$A_\mu$ ($p_\mathrm{T}^\mu> $20\gevc) } & \multicolumn{3}{|c|}{$A_\mu$ ($p_\mathrm{T}^\mu> $25\gevc) } &\multicolumn{3}{|c}{$A_\mu$ ($p_\mathrm{T}^\mu> $30\gevc) }\\ \hline
$2.0 - 2.5$ & $0.275  $&$ \pm  0.014 $&$ \pm  0.003$&  $0.256$&$ \pm  0.015$&$  \pm  0.002$ & $0.238  $&$ \pm  0.018 $&$ \pm  0.002$\\
$2.5 - 3.0$ & $0.178  $&$ \pm  0.013 $&$ \pm  0.002$& $0.195$&$  \pm  0.015$&$  \pm  0.001$ & $0.219  $&$ \pm  0.017 $&$ \pm  0.001$\\
$3.0 - 3.5$ & $0.024  $&$ \pm  0.016 $&$ \pm  0.009$& $0.054$&$  \pm  0.018$&$  \pm  0.003$ & $0.112  $&$ \pm  0.022 $&$ \pm  0.002$\\
$3.5 - 4.0$ & $-0.247 $&$ \pm  0.022 $&$ \pm  0.011$& $-0.203$&$  \pm  0.027$&$  \pm  0.005$ & $-0.124 $&$ \pm  0.035 $&$ \pm  0.003$\\
$4.0 - 4.5$ & $-0.493 $&$ \pm  0.058 $&$ \pm  0.051$& $-0.413$&$  \pm  0.081$&$  \pm  0.016$ & $-0.353 $&$ \pm  0.122 $&$ \pm  0.008$\\   

\end{tabular}
}

\end{center}
\vspace{1cm}
\label{tab:asy2}
\caption{$W$ cross-section ratio, $R_W=\sigmawp/\sigmawm$,  in bins of muon
 pseudorapidity. The first error is statistical and the second systematic.  The effect of FSR is at the level of 
 $10^{-4}$ and is not listed.
}
\begin{center}
\begin{tabular}{c|cll}
$\eta^{\mu}$ & \multicolumn{3}{|c}{$R_W$} \\ \hline
$2.0 - 2.5$ &  $1.76$&$ \pm  0.05$&$  \pm  0.01$ \\
$2.5 - 3.0$ & $1.43$&$  \pm  0.04$&$  \pm  0.01$ \\
$3.0 - 3.5$ & $1.05$&$  \pm  0.03$&$  \pm  0.02$ \\
$3.5 - 4.0$ & $0.60$&$  \pm  0.03$&$  \pm  0.01$ \\
$4.0 - 4.5$ & $0.34$&$  \pm  0.05$&$  \pm  0.05$ \\  
\end{tabular}

\end{center}
\label{tab:wratio}
\end{table}


\end{document}